\documentclass[journal]{IEEEtran}
\ifCLASSOPTIONcompsoc
  \usepackage[nocompress]{cite}
\else
  \usepackage{cite}
\fi
\newtheorem{definition}{Definition}[section]
\newtheorem{theorem}{Theorem}[section]
\usepackage{comment}
\usepackage{amsmath,amssymb,amsfonts}
\usepackage{algorithmic}
\usepackage{graphicx}
\usepackage{fancybox}
\usepackage{xcolor}
\usepackage{multirow}

\usepackage{algorithm, algorithmic}
\usepackage{enumitem}
\usepackage{url}

\DeclareMathOperator*{\argmax}{argmax}

\begin{document}
%
\title{Hide and Seek: Outwitting Community \\Detection Algorithms}

\author{\IEEEauthorblockN{Shravika Mittal\textsuperscript{$\star$},
        Debarka Sengupta\textsuperscript{$\star$,$\dagger$}, Tanmoy Chakraborty\textsuperscript{$\star$}}\\
\IEEEauthorblockA{\textit{\textsuperscript{$\star$}Dept. of CSE, \textsuperscript{$\dagger$}Dept. of Comp. Biology, IIIT-Delhi, India}\\
\{shravika16093, debarka, tanmoy\}@iiitd.ac.in}
        }
\IEEEtitleabstractindextext{%
\begin{abstract}
Community affiliation of a node plays an important role in determining its contextual position in the network, which may raise privacy concerns when a sensitive node wants to hide its identity in a network. Oftentimes, a  target community seeks to protect itself from adversaries so that its constituent members remain hidden inside the network. The current study focuses on hiding such sensitive communities so that community affiliation of the targeted nodes can be concealed. This leads to the problem of {\em community deception} which investigates the avenues of minimally rewiring nodes in a network so that a given target community maximally hides from a community detection algorithm.
We formalize the problem of community deception and introduce NEURAL, a novel method that greedily optimizes a {\em node-centric objective function} to determine the rewiring strategy. Theoretical settings pose a restriction on the number of strategies that can be employed to optimize the objective function, which in turn reduces the overhead of choosing the best strategy from multiple options. We also show that our objective function is {\em submodular} and {\em monotone}. When tested on both synthetic and 7 real-world networks, NEURAL is able to deceive 6 widely used community detection algorithms. We benchmark its performance with respect to 4 state-of-the-art methods on 4 evaluation metrics. Additionally, our qualitative analysis on 3 other attributed
real-world networks reveals that NEURAL, quite strikingly, captures important meta-information about edges that otherwise could not be inferred by observing only their topological structures.
\end{abstract}

\begin{IEEEkeywords}
Community detection, community hiding, permanence, complex networks
\end{IEEEkeywords}}

\maketitle

\IEEEdisplaynontitleabstractindextext

%
\IEEEpeerreviewmaketitle

\section{Introduction}\label{sec:introduction}
\IEEEPARstart{D}{etecting} communities from large networks has remained as one of the major research problems in the last two decades. Different heuristics, metrics, and optimization techniques have been proposed to detect communities from multiple types of networks \cite{fortunato2016community}. However, of late, limited efforts have been visible to understand how easily a community detection algorithm can be deceived by minimal rewiring of nodes. 

In this paper, we ask a fundamental question: \textbf{\textit{How do we hide a target community from being exposed to a community detection algorithm, assuming limited rewiring operations are allowed?}} In other words, can a node or a community disguise its positioning in the network in order to escape detection \cite{Waniek_2018}? We call this problem \textbf{Hide and Seek Community (HSC)}. Answering this question matters since it helps the social network users in hiding their identity from online surveillance\footnote{Mislove et al. \cite{mislove2010you} showed how by breaking down Facebook user network and attributes of certain users, it is possible to gather private data about other Facebook users.}  \cite{8999786}. It also helps law-enforcement organizations identify criminal acts deceiving online identity \cite{8118127}. This may also be useful for counter-terrorism units in order to deploy spies into a terrorist network. The solution of the current problem would help the spies determine who they should start a new friendship with (edge addition) or which existing friendship they should try to break (edge deletion) to conceal their community identity. However, one may argue that the same method can be misused by the adversaries.  Nonetheless, we believe that our investigation brings issues to light for the plan of novel community detection methods vigorous to deception strategies.


To date, the fundamental question stated above has got very little attention as most of the focus has been concentrated towards building efficient algorithms for community detection. Nagaraja \cite{inproceedings} made a pioneering attempt to examine the degree of network information required by an attacker to infer the community membership information. Recently, Waniek et al. \cite{Waniek_2018} proposed a heuristic-based solution to evade network centrality analysis.  Fionda and Pirr\`{o} \cite{8118127} proposed a novel metric and greedily optimized it to hide the members of a target community from being detected by the community detection algorithms. Liu et al. \cite{NIPS2019_9453} proposed an approach to maximally hide the {\em entire} community structure (as opposed to a target community)   with a minimum rewiring of the network structure. 

\begin{figure}[!t]
  \centering
  \includegraphics[width=\columnwidth]{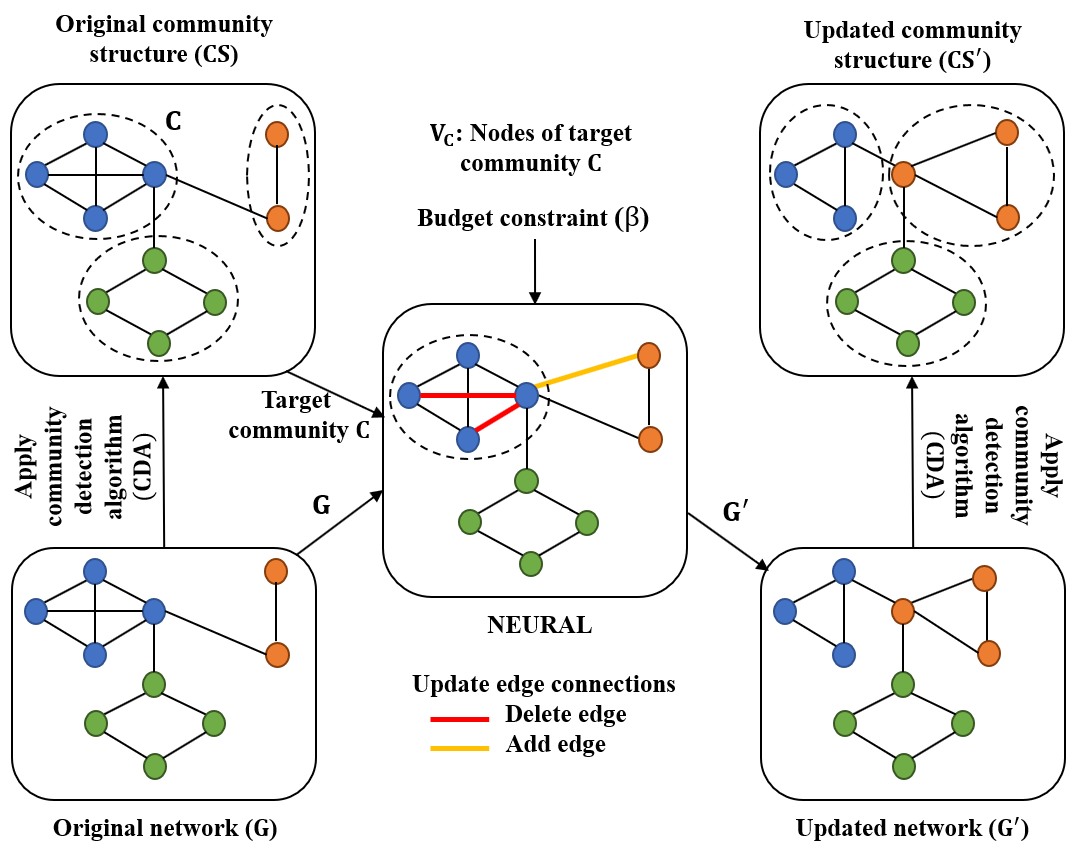}
  \caption{Flow diagram showing the procedure of NEURAL.}
  \label{fig:diagram}
\end{figure}


Here, we pose the HSC problem as a constrained optimization problem. The objective function is designed based on {\bf Permanence} \cite{Chakraborty:2014:PVN:2623330.2623707}, a node-centric metric we proposed previously,  which has been proved to be highly effective in detecting the entire community structure of a network. Permanence, being a local metric, uses limited information of a node to determine its community membership. We theoretically prove that only two types of edge update operations  (inter-community edge addition and intra-community edge deletion) are useful for rewiring nodes to optimize our proposed objective function. We further show that the objective function is {\em submodular} and {\em monotone} w.r.t. the required edge updates.  Therefore, we propose {\bf NEURAL} ({\bf Ne}twork deception {\bf u}sing pe{\bf r}m{\bf a}nence {\bf l}oss), a greedy optimization algorithm to optimize the objective function. Given a network $G$, its community structure {\em CS} obtained from a community detection algorithm {\em CDA}, and a target community $C$ whose constituent nodes $V_C$ need to be concealed,  NEURAL rewires nodes within the rewiring budget $\beta$ in such a way that {\em CDA} is unable to identify the original community affiliation of $V_C$ (Fig. \ref{fig:diagram} shows a flow diagram). 

Extensive experiments are conducted on both synthetic and 7 real-world networks. Six widely used community detection algorithms are considered for deception. We compare NEURAL with 4 state-of-the-art community deception methods. The performance is measured based on 4 evaluation metrics (two of them are proposed by us). Our quantitative analysis shows that  NEURAL significantly outperforms others  across all the datasets and all the evaluation metrics. 

We further conduct a detailed qualitative analysis to explain the physical significance of edges selected by the deception methods on three attributed real-world networks -- citation network, terrorist network, and breast cancer network. Surprisingly, we observe that NEURAL is able to capture important meta-information of edges that otherwise could not be inferred just by observing the topological structure of networks.   

In short, our major contributions are four-fold:
\begin{itemize}
    \item {\bf Novel objective function:} Our proposed objective function is novel which considers minimum information of nodes for network rewiring.  
    \item {\bf Novel algorithm:} We propose NEURAL, a novel greedy optimization algorithm for community deception.
    \item {\bf Quantitative evaluation:} We perform an extensive evaluation on multiple datasets and show that NEURAL outsmarts existing approaches for hiding the target community within the specific budget.
    \item {\bf Qualitative evaluation:} We further interpret the edges selected for node rewiring by the deception methods and show that NEURAL captures important meta-information of edges in three real-world networks. 
    \end{itemize}

{\bf Reproducibility:} The codes and datasets are available at: \url{https://github.com/mittalshravika/HideAndSeek-NEURAL}. 

\section{Related work}

\begin{table}[!t]\label{tab:2}
\centering
\caption{Comparison of NEURAL with existing methods.\\ Abbreviations: $\mathbf{E +,-}$: edge addition and deletion, NC: node centrality, $\mathbf{E\rightarrow C}$: edges connected with the nodes in the target community $\mathbf{C}$, QA: qualitative analysis.}
\begin{tabular}{ccccc}
\hline
Method & Metric & Strategy & Knowledge & QA \\
\hline
Nagaraja \cite{inproceedings} & Modularity & $E\ +$ & NC & No \\
DICE \cite{Waniek_2018}  & Modularity & $E +,-$ & $E \rightarrow C$ & No \\
SADDEN \cite{8118127}  & Safeness & $E +,-$ & $E \rightarrow C$ & No \\
\hline
{\bf NEURAL} & Permanence & $E +,-$ & $E \rightarrow C$ & Yes \\
\hline
\end{tabular}
\end{table}

{\bf Community Detection:} There has been a plethora of research in the detection of communities from a given network. These include traditional clustering based algorithms such as hierarchical clustering, partitional clustering and spectral clustering, which group nodes together based on a similarity metric \cite{8566218}. Another class of community detection algorithms revolves around the optimization of metrics that define the quality of a network partition, such as modularity \cite{Newman_2004b, 6785984}, conductance \cite{doi:10.1080/15427951.2009.10129177}, cut-ratio \cite{10.1145/1772690.1772755}, etc. 
Few other methods are based on random walks \cite{10.1007/11569596_31}, information theory \cite{Rosvall1118,Rosvall7327}, and spectral algorithms \cite{Donetti_2004,RePEc:eee:phsmap:v:352:y:2005:i:2:p:669-676}. 
Algorithms that detect overlapping communities have also been proposed \cite{Palla2005,8360525}. 
A detailed study of community detection algorithms can be found in \cite{Fortunato_2010,chakraborty2017metrics}. 

{\bf Community Deception:} Another area of interest that has started revolving very recently is {\em community deception} i.e., hiding a target community or the entire community structure  from getting exposed to community detection algorithms. 
Nagaraja \cite{inproceedings} proposed a counter detection method for hiding a community by adding edges under a certain budget. The endpoints of edges to be added are chosen using vertex centrality measures (degree centrality, eigenvector centrality, and random initialization). 
Waniek et al. \cite{Waniek_2018} proposed DICE, an algorithm that deletes intra-community edges (\textit{disconnect internal}) and adds inter-community edges (\textit{connect external}), inspired by the functioning of modularity. The authors also devised a metric to quantify the concealment of a target community in the network. Fionda and Pirr\`{o} \cite{8118127} referred to the problem of hiding a community as \textit{community deception}. They devised a greedy optimization algorithm (dubbed SADDEN henceforth) to hide a target community based on \textit{safeness gain}, a new metric that they proposed to quantify how safe a node is under adversarial attack. SADDEN requires the knowledge of the local community rather than knowing the entire community structure of the network to deceive community detection algorithms. Along with this, the authors proposed a metric, called \textit{deception score} to quantify the effect of the community deception algorithm on the network.  They also showed that their method outperforms modularity-based approaches. Recently, Liu et al. \cite{NIPS2019_9453} extended the problem of hiding a target community to hiding the {\em entire community structure}. They  proposed an algorithm for \textit{community structure deception} based on information theory using network entropy minimization.

We consider all the methods mentioned above (Nagaraja, DICE, SADDEN)  as baselines\footnote{To our knowledge, these are the only existing methods which attempted to solve the HSC problem.} along with a random edge rewiring method, except Liu et al. \cite{NIPS2019_9453} as this method focuses on the deception of the {\em entire community structure} (instead of a {\em single target community}); moreover, the metric used in their method (community-based structural entropy) requires entire community information.

{\bf How NEURAL is different from others?}
Table \ref{tab:2} summarizes how NEURAL is different from the existing methods for community deception. NEURAL uses Permanence  as a metric to determine how to update a given network efficiently in order to hide the target community. We  perform comprehensive evaluation on both synthetic and real-world networks using four different evaluation metrics. We further perform a qualitative analysis on 3 attributed networks to understand the significance of the selected edges. 

\section{Problem Formulation}
\subsection{Preliminaries}
A network $G = (V, E)$ is defined as an undirected graph with $V$ as the set of vertices and $E$ as the set of edges. After applying a community detection algorithm on $G$, we get $CS = (C_{1}, C_{2},...C_{k})$ as the community structure. We only consider communities that are non-overlapping. For community $C\in CS$, an intra-community edge $\langle u, v \rangle$ is defined such that $u, v$ $ \in $ $C$, and an inter-community edge $\langle u, v \rangle$ is defined such that $u$ $ \in $ $C$ and $v \in C'$ where $C \cap C' = \phi$. $E_{\mathrm{intra}}(C)$ ({\em resp.} $E_{\mathrm{inter}}(C)$) denotes the set of intra- ({\em resp.} inter-) community edges corresponding to $C$.  

\subsection{Hide and Seek Community (HSC)}
Our primary goal is to come up with an algorithm that, with minimum edge rewiring, is able to hide a given target community $C$ from a community detection method. In other words, the actual community membership information of nodes inside $C$ should not be revealed by the community detection method.  This is done by rearranging the structure of the network using a certain number ($\beta$) of  edge updates (which we  call {\em budget} for network rewiring). We also assume that each edge update operation will incur a unit cost. 
One approach would be to search through the entire space for possible edge updates exhaustively and select the ones that are able to hide the target community $C$ the most. However, searching through this huge space of all the possible combinations of edge updates would become computationally expensive in case of large networks. Along with this, such an exhaustive technique would require the knowledge of the entire network and may also depend on the type of community detection algorithm that we intend to fool.

To avoid this, we introduce the problem, called \textit{Hide and Seek Community} to camouflage a target community $C$ from a community detection method. 

\begin{definition}
{\bf (Hide and Seek Community)} For a network $G = (V, E)$, the problem of \textit{Hide and Seek Community} (HSC) is to hide a target community $C$ with the help of network edge updates constrained by a parameter $\beta$. It can be posed as a constrained optimization problem as follows:
\begin{equation}\label{eq:optfun}
    \begin{split}
    & \argmax_{E'(C)} \mathcal{F}(C, E(C), \beta, E'(C))
\end{split}
\end{equation}
where, $E(C) = E_{\mathrm{intra}}(C) \cup E_{\mathrm{inter}}(C)$,
$E'(C) = (E(C) \cup E_{\mathrm{add}}) \setminus E_{\mathrm{del}}$, and $E_{\mathrm{add}}$ ({\em resp.} $E_{\mathrm{del}}$) indicates the set of edges to be added ({\em resp.} deleted) to hide $C$ such that $\vert E_{\mathrm{add}} \vert + \vert E_{\mathrm{del}} \vert \leq \beta$.
\end{definition}


\section{Methodology}
We consider {\em Permanence} \cite{Chakraborty:2014:PVN:2623330.2623707,chakraborty2016permanence}, a node-centric metric\footnote{Permanence can be also computed for an entire network.} to design the objective function $\mathcal{F}$ in  \eqref{eq:optfun}. We theoretically show that limited edge update operations are required to maximize the {\em Permanence loss} (our objective function). We also show that  Permanence loss is submodular and monotone w.r.t. each of the edge update operations.  Therefore, we propose NEURAL, a greedy algorithm that makes use of  Permanence loss in order to hide a target community $C$. This section first briefly describes Permanence, followed by the greedy strategy used in  NEURAL.

\subsection{Permanence}
\begin{figure}[!t]
  \centering
  \includegraphics[scale=0.8]{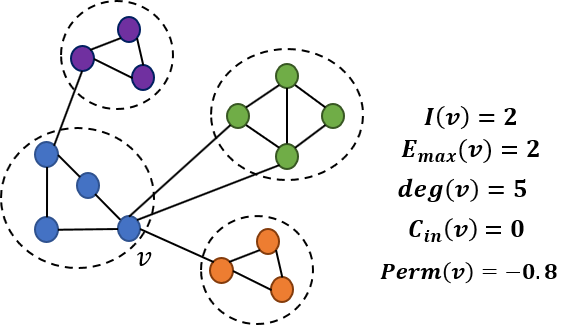}
  \caption{A toy example demonstrating the calculation of Permanence for a node, given the network and the community structure.}
  \label{fig:2}
\end{figure}

Chakraborty et al. proposed Permanence \cite{Chakraborty:2014:PVN:2623330.2623707,chakraborty2016permanence}, a vertex-centric metric that quantifies the containment of a node $v$ in a network community $C$. The formulation of Permanence is based on three factors - (i) the internal pull $I(v)$, denoted by the internal connections of a node $v$  within its own community, (ii) maximum external pull $E_{max}(v)$, denoted by the maximum connections of $v$ to its neighboring communities, and (iii) internal clustering coefficient of $v$, $C_{in}(v)$, denoted by the fraction of actual and possible number of edges among the internal neighbors of $v$. The above three factors are then suitably combined to obtain the Permanence of $v$ as,
\begin{equation}\label{eq:1}
    \begin{split}
    Perm(v,G) = \frac{I(v)}{E_{max}(v)}\times\frac{1}{deg(v)} - \large(1 - C_{in}(v)\large)\end{split}
\end{equation}
Fig. \ref{fig:2} shows a toy example to calculate the Permanence value of a node. 

This metric indicates that a vertex would remain in its own community as long as its internal pull is greater than the external pull or  its internal neighbors are densely connected to each other, hence forming a near clique.  
The Permanence for a network \textit{G} is then defined as $
    Perm(G) = \frac{\sum_{v \in V} Perm(v)}{\vert V \vert}$.
The reasons behind choosing Permanence instead of other community scoring metrics such as (local) modularity \cite{newman2006modularity,muff2005local}, conductance, cut-ratio \cite{fortunato2016community} are two-fold: (i) Permanence is a vertex-centric local metric which would enable us to update edges incrementally in order to change the network structure without looking into the entire network structure, and (ii) Permanence has been shown to be superior to other local and global scoring metrics for community detection \cite{chakraborty2017metrics}.

\subsection{Proposed Objective Function: Permanence Loss}
Our proposed community deception method NEURAL (discussed in Section \ref{sec:algo}) aims to reduce  Permanence of the network for a target community $C$ to be hidden from community detection algorithms. We propose to do so because reducing Permanence of a vertex would disrupt its containment in the original community, changing the community structure of the network, making it difficult for detection algorithms to identify the original communities. 
We search for edge updates (addition/deletion of edges) by maximizing the {\bf Permanence loss} at every iteration,  defined as,
\begin{equation}
    \mathcal{P}_{l} = Perm(G) - Perm(G')
\end{equation}
where $G$ represents the original network, and $G'$ represents the modified network after updating edges (see Fig. \ref{fig:diagram}) w.r.t. the target community $C$, elaborated in Sections \ref{sec:edge_update} and \ref{sec:algo}.

In Section \ref{sec:edge_update}, we will show that Permanence loss will be affected (positively) only due to the (i) intra-community edge deletion, and (ii) inter-community edge addition. Readers are encouraged to see supplementary where we show that {\bf Permanence loss is submodular and monotone w.r.t. each of the  edge updates stated above}. 

\subsection{Edge Updates}\label{sec:edge_update}
In this section, we describe four possible edge update operations  to maximise Permanence loss $\mathcal{P}_{l}$ in  NEURAL --- inter- and intra-community edge deletion, and inter- and intra-community edge addition.

\subsubsection{Inter-community Edge Deletion}

{\bf         
\begin{theorem} \label{theorem:1}
        Deleting an inter-community edge $\langle u, v \rangle$ where $u  \in C$ and $v \in C'$ such that $C \cap C' = \phi$, does not result in  Permanence loss.
\end{theorem}}
\begin{IEEEproof}
In this proof, we show that deleting an inter-community edge does not amount to Permanence loss. An inter-community edge deletion just affects the Permanence measure for $u$ and $v$. We will only show the change in Permanence for node $u$ (same applies to $v$). There can be two cases:\\
        (i) $\mathbf{E_{max}(u)}$ {\bf does not change after edge deletion:}
                In this case, we assume that the maximum external connections for node $u$ remain the same after deleting $\langle u, v \rangle$. Deleting $\langle u, v \rangle$ would not change $C_{in}(u)$. It would only decrease its degree by 1. Therefore, for Permanence loss, we need to see whether $\mathcal{P}_{l} = Perm(u, G) - Perm(u, G') \geq 0$. This reduces to, 
                \begin{equation*}\small
                    \mathcal{P}_{l} = \frac{I(u)}{E_{max}(u)}\times\left[\frac{1}{deg(u)} - \frac{1}{deg(u) - 1}\right] < 0
                \end{equation*}
                Therefore, no Permanence loss is possible in this case.\\
            (ii) $\mathbf{E_{max}(u)}$ \textbf{changes after edge deletion:}
                In this case, we assume that the deletion of edge $\langle u, v \rangle$ affects the maximum external connections of node $u$. This is the case where $C'$ is the only community that has the maximum external pull for node $u$. As a result, along with degree, $E_{max}(u)$ would also decrease by 1. It would not change $C_{in}(u)$. Therefore, for Permanence loss, we need to see whether $\mathcal{P}_{l} = Perm(u, G) - Perm(u, G') \geq 0$. This reduces to,
                \begin{equation*}\small
                \begin{split}
                \mathcal{P}_{l} & = I(u)\left[\frac{1}{E_{max}(u)\times deg(u)} - \frac{1}{(E_{max}(u) - 1)\times(deg(u) - 1)}\right]\\
                & = I(u)\left[\frac{1 - E_{max}(u) - deg(u)}{E_{max}(u)\times deg(u)\times(E_{max}(u) - 1)\times(deg(u) - 1)}\right]\\
                & < 0 \text{ ($E_{max}(u) \geq 1$ and $deg(u) \geq 1$ because of edge $\langle u, v \rangle$)}
                \end{split}
                \end{equation*}
            
                Therefore, no Permanence loss is possible in the case of deleting an inter-community edge.
            
        \end{IEEEproof}

    \subsubsection{Intra-community Edge Deletion}
        {\bf 
        \begin{theorem}\label{theorem:2}
        Deleting an intra-community edge $\langle u, v \rangle$ where $u, v \in C$, always results in Permanence loss.
        \end{theorem}}
        \begin{IEEEproof}
            Here we show that deleting an intra-community edge always results in Permanence loss. We will only show the change in Permanence for node $u$ (same applies to $v$). Such an edge update would decrease the internal degree and degree of node $u$ by 1. It would not affect $E_{max}(u)$ (no external connections are being changed). We narrow our search space such that, $C_{in}(u)$ decreases after the deletion of $\langle u, v \rangle$. Therefore, for  Permanence loss, we need to see whether $\mathcal{P}_{l} = Perm(u, G) - Perm(u, G') \geq 0$. This reduces to,
            \begin{equation}\label{eq:del}
            \begin{split}
            \mathcal{P}_{l} & = \frac{1}{E_{max}(u)}\left[\frac{I(u)}{deg(u)} - \frac{I(u) - 1}{deg(u) - 1}\right]\\
            & = \frac{1}{E_{max}(u)}\left[\frac{deg(u) - I(u)}{deg(u)\times(deg(u) - 1)}\right]\\
            & \geq 0 \text{ (as $deg(u) \geq I(u)$)}
            \end{split}
            \end{equation}
            Therefore, deleting an intra-community edge $\langle u, v \rangle$ would bring in  Permanence loss in terms of nodes $u$ and $v$. 
            
            The intra-community edge deletion would also affect the Permanence measure for nodes that have both $u$ and $v$ as their neighbors. If so, it would result in a change in their internal clustering coefficient value with all the other factors unchanged. For  Permanence loss due to such a node $w$, we need to see whether $\mathcal{P}_{l} = Perm(w, G) - Perm(w, G') \geq 0$. This reduces to,
            \begin{equation*}\small
            \begin{split}
            \mathcal{P}_{l} & = (1 - C_{in}'(w)) - (1 - C_{in}(w))
             = C_{in}(w) - C_{in}'(w)
            \end{split}
            \end{equation*}
            where $C_{in}'(w)$ represents the updated internal clustering coefficient of $w$. In the above equation, $\mathcal{P}_{l} > 0$ since $C_{in}(w) > C_{in}'(w)$. For node \textit{w}, the number of neighbors is intact, but the edges between its neighbors get reduced by 1 after $\langle u, v \rangle$ is deleted. As a result, the internal clustering coefficient reduces, again resulting in Permanence loss. 
            Therefore, deleting $\langle u, v \rangle$ would also bring in Permanence loss in terms of their common neighbors.
            \end{IEEEproof}

 
    \subsubsection{Inter-community Edge Addition}
       {\bf 
        \begin{theorem}\label{theorem:3}
        Adding an inter-community edge $\langle u, v \rangle$ where $u \in C$ and $v \in C'$, such that $C \cap C' = \phi$, always results in  Permanence loss. The loss is more if $C'$ is the community that provides the maximum external pull for node $u$.
        \end{theorem}}
        \begin{IEEEproof}
        In this proof, we show that adding an inter-community edge always causes  Permanence loss. An inter-community edge addition just affects Permanence for nodes $u$ and $v$. We will only show the change in Permanence for node $u$ (same applies to $v$). There can be two cases:\\
        (i) $\mathbf{E_{max}(u)}$ \textbf{does not change after edge addition:}
            In this case, we assume that the maximum external connections for node $u$ remain the same after adding $\langle u, v \rangle$. Adding $\langle u, v \rangle$ would have no effect on $C_{in}(u)$. It would only increase its degree by $1$. So, for  Permanence loss, we need to see whether $\mathcal{P}_{l} = Perm(u, G) - Perm(u, G') \geq 0$. This reduces to,
            \begin{equation}\label{eq:4}\small
                \mathcal{P}_{l} = \frac{I(u)}{E_{max}(u)}\times\left[\frac{1}{deg(u)} - \frac{1}{deg(u) + 1}\right] > 0
            \end{equation}
            Therefore, there is a  Permanence loss in the case of adding an inter-community edge such that $E_{max}(u)$ does not change after edge addition.\\
        (ii) $\mathbf{E_{max}(u)}$ \textbf{changes after edge addition:}
            In this case, we assume that the addition of edge $\langle u, v \rangle$ affects the maximum external connections of node $u$. This is the case where $C'$ is the community that has the maximum external pull for node $u$. As a result, along with the degree, $E_{max}(u)$ would also increase by 1. It would not change $C_{in}(u)$. Therefore, for  Permanence loss, we need to see whether $\mathcal{P}_{l} = Perm(u, G) - Perm(u, G') \geq 0$. This reduces to,
            \begin{equation}\label{eq:5}\small
            \begin{split}
            \mathcal{P}_{l} & = I(u)\left[\frac{1}{E_{max}(u)\times deg(u)} - \frac{1}{(E_{max}(u) + 1)\times(deg(u) + 1)}\right]\\
            & = I(u)\left[\frac{1 + E_{max}(u) + deg(u)}{E_{max}(u)\times deg(u)\times(E_{max}(u) + 1)\times(deg(u) + 1)}\right]\\
            & > 0 
            \end{split}
            \end{equation}
            Therefore, there is  Permanence loss in the case of adding an inter-community edge such that $E_{max}(u)$ changes after edge addition.
            \end{IEEEproof}
       {\bf      
        \begin{theorem}\label{theorem:5}
        The Permanence loss is more in case of \eqref{eq:5} (i.e., an edge added to the neighboring community from where $u$ experiences the maximum external pull) as compared to  \eqref{eq:4}.
        \end{theorem}}
         
        \begin{IEEEproof}
        Taking Permanence loss in \eqref{eq:4} and \eqref{eq:5}, we get, 
        \begin{equation*}\small
        \begin{split}
        & I(u)\left[\frac{1 + E_{max}(u) + deg(u)}{E_{max}(u)\times deg(u)\times(E_{max}(u) + 1)\times(deg(u) + 1)}\right]\\
        & \geq \frac{I(u)}{E_{max}(u)}\times\left[\frac{1}{deg(u)\times(deg(u) + 1)}\right]\\
        & \Rightarrow deg(u) \geq 0 \text{ which is true.}
        \end{split}
        \end{equation*}
        \end{IEEEproof}

        \subsubsection{Intra-community Edge Addition}
        
\begin{figure}[!t]
  \centering
  \includegraphics[scale = 0.8]{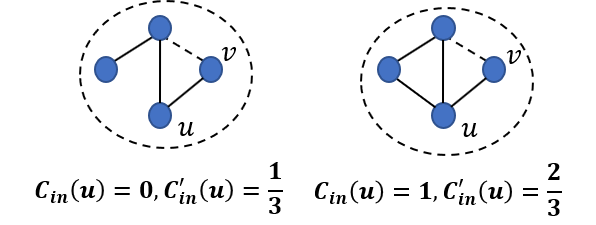}
  \vspace{-3mm}
  \caption{An example to demonstrate that Permanence loss in terms of cohesiveness of internal neighbors of a node \textit{u} may not always be positive.}
  \label{fig_permloss}
  \vspace{-5mm}
\end{figure}

       {\bf  
        \begin{theorem}\label{theorem:4}
        Adding an intra-community edge $\langle u, v \rangle$ where $u, v \in C$ does not always ensure a loss in Permanence.
        \end{theorem}}
        \begin{IEEEproof}
            Here we show that adding an intra-community edge does not always result in Permanence loss. We will only show the change in Permanence for node $u$ (same applies to $v$).
            
            For this, we consider two parts of Permanence  separately - (i) ratio of internal-external pull, denoted by $Perm(G)_{1}$, and (ii) cohesiveness of internal neighbors, denoted by $Perm(G)_{2}$. \\
             \textbf{(i) Impact on the ratio of internal-external pull:}
                In this, we consider the effect of adding an intra-community edge $\langle u, v \rangle$ on the internal-external pull factor of  Permanence. This update increases the internal degree and degree for node $u$ by 1. It has no effect on the maximum external connections $E_{max}(u)$. Therefore, for Permanence loss we need to see whether $\mathcal{P}_{l1} = Perm(u, G)_{1} - Perm(u, G')_{1} \geq 0$. This reduces to, 
                \begin{equation*}\small
                \begin{split}
                \mathcal{P}_{l1} & = \frac{1}{E_{max}(u)}\left[\frac{I(u)}{deg(u)} - \frac{I(u) + 1}{deg(u) + 1}\right]\\
                & = \frac{1}{E_{max}(u)}\left[\frac{I(u) - deg(u)}{deg(u)\times(deg(u) + 1)}\right]\\
                & \leq 0 \text{ (as $I(u) \leq deg(u)$)}
                \end{split}
                \end{equation*}
                Therefore, there is no Permanence loss w.r.t. the internal-external pull (first part of \eqref{eq:1}).\\
                \textbf{(ii) Impact on cohesiveness of internal neighbors:}
                In this, we consider the effect of adding an intra-community edge $\langle u, v \rangle$ on the cohesiveness of internal neighbors. Therefore, for  Permanence loss, we need to see whether $\mathcal{P}_{l2} = Perm(u, G)_{2} - Perm(u, G')_{2} \geq 0$. This reduces to,
                \begin{equation*}\small
                \begin{split}
                \mathcal{P}_{l2} & = (1 - C_{in}'(u)) - (1 - C_{in}(u)) = C_{in}(u) - C_{in}'(u)
                \end{split}
                \end{equation*}
                where $C_{in}'(u)$ represents the updated internal clustering coefficient of  $u$ in $G$.  $\mathcal{P}_{l2}$ can be positive or negative depending on how the connections between internal neighbors of $u$ change after introducing its new neighbor $v$. This is shown using a toy example in Fig. \ref{fig_permloss}. It can be seen that in the 1st case, $\mathcal{P}_{l2} < 0$, while in the second case, $\mathcal{P}_{l2} > 0$. 
             
        By combining   (i) and (ii), we conclude that intra-community edge addition does not always ensure  Permanence loss.
    \end{IEEEproof}

\subsection{Proposed Algorithm: NEURAL}\label{sec:algo}
Since our objective function is submodular and monotone w.r.t. the possible edge updates that affect Permanence loss positively, we propose NEURAL, a greedy algorithm that maximizes  Permanence loss to rewire nodes within a given budget in order to hide the target community. 

NEURAL makes the use of certain edge updates discussed in the previous section to rewire the network structure such that the community detection algorithms are not able to detect a target community \textit{C}. Along with the network, it takes as input $\beta$, indicating the budget or the maximum number of edge updates that are allowed. 
The pseudo-code of NEURAL is shown in Algorithm 1 (flow diagram in Fig. \ref{fig:diagram}). 
At every iteration, it considers an edge update which contributes towards the maximum loss in Permanence for the network, hence greedily updating the original network.
For an edge addition, we only consider adding inter-community edges following Theorems \ref{theorem:3} and  \ref{theorem:4} as it has been shown that adding an intra-community edge does not guarantee a loss in Permanence in all cases (lines \ref{line:add_start}-\ref{line:add_end} of Algorithm \ref{alg:1}). In the case of edge deletion, we only consider deleting intra-community edges following Theorems \ref{theorem:1} and \ref{theorem:2} (lines \ref{line:del_start}-\ref{line:del_end} of Algorithm \ref{alg:1}). Deleting an inter-community edge does not result in Permanence loss in any case; hence it is not a favorable update. Since NEURAL follows a greedy strategy, for the addition of all the competing inter-community edges, the one which has the highest Permanence loss for the network is considered. The same approach is followed for selecting the best intra-community edge for deletion. In the end, a choice between the best inter-community edge to be added and the best intra-community edge to be deleted is made based on which one contributes more to network Permanence loss (lines \ref{line:choic_start}-\ref{line:choic_end} of Algorithm \ref{alg:1}).

Note that for computing the best network update at every iteration, we only need node information for a subset of all the nodes present in the network which reduces the amount of network information being used. 

\subsection{Time Complexity of NEURAL}
The time complexity of NEURAL is $\mathcal{O}(\vert V_{C} \vert + \vert E_{C} \vert)$, where $\vert V_{C} \vert$ and $\vert E_{C} \vert$ represent the number of nodes and edges (both intra-community and inter-community) in the target community \textit{C}, respectively. This is because, in order to search for edge updates that best contribute towards the Permanence loss for hiding $C$, we only need to go through the nodes and edge connections in the target community as shown in Section \ref{sec:edge_update}. Information about the rest of the network is not required. We explore the running time complexity of NEURAL further in supplementary.

\begin{algorithm}[!ht]\small
\caption{NEURAL: Network Deception using Permanence Loss}
\label{alg:1}
\begin{algorithmic}[1]
\REQUIRE (i) Network $G$, (ii) target community $C$, (iii) budget $\beta$
\ENSURE Updated Network $G'$ 
\STATE $\mathrm{\mathcal{P}_{l, add} = 0}$ \label{line:1}
\STATE $\mathrm{\mathcal{P}_{l, del} = 0}$ \label{line:2}
    \WHILE {$\beta > 0$}
        \STATE $\mathrm{add_{u}, maxComm_{u} = getBestNodeForAddition(C)}$ \eqref{eq:5} \label{line:add_start}
        \STATE $\mathrm{add_{v} = getBestExternalNodeForAddition(maxComm_{u})}$
        \STATE $\mathrm{\mathcal{P}_{l, add} = getEdgeAdditionLoss(add_{u}, add_{v})}$ \label{line:add_end}
        \STATE $\mathrm{intraEdge \leftarrow \mathrm{getConnectingEdges(C)}}$ \label{line:del_start}
        \STATE $\mathrm{del_{u}, del_{v} = getBestEdgeForDeletion(intraEdge)}$ \eqref{eq:del}
        \STATE $\mathrm{\mathcal{P}_{l, del} = getEdgeDeletionLoss(del_{u}, del_{v}, C)}$ \label{line:del_end}
        
        \IF{$\mathrm{\mathcal{P}_{l, add} \geq \mathcal{P}_{l, del}}$ and  $\mathrm{\mathcal{P}_{l, add} > 0}$} \label{line:choic_start}
        \STATE $\mathrm{G \leftarrow (V, E \cup \{add\textunderscore u, add\textunderscore v\})}$
        \ELSIF{$\mathrm{\mathcal{P}_{l, del} > 0}$}
        \STATE $\mathrm{G \leftarrow (V, E \backslash \{del\textunderscore u, del\textunderscore v\})}$ \label{line:choic_end}
        \ENDIF
        \STATE $\beta = \beta - 1$
    \ENDWHILE 
    \RETURN G
\end{algorithmic}
\end{algorithm}

\vspace{-7mm}
\section{Experimental Setup}
In this section, we start by briefly describing the datasets, baseline methods, community detection methods we considered for deception, and the evaluation metrics. We then elaborate on the experimental results and the case studies. 

\subsection{Synthetic and Real-world Networks}
We conduct experiments on two types of networks:\\
\noindent\textbf{(i) Synthetic networks:} We use  LFR Benchmark \cite{Lancichinetti2008Benchmark} and vary the following parameters to generate synthetic networks: $N$, number of nodes and $\mu$, the ratio of external connections of a node to degree. The other parameters are set to default as mentioned in the original implementation.
Unless otherwise stated, we consider the following setting to generate the default synthetic network: $N=10,000$, $\mu=0.4$ (as suggested in \cite{Chakraborty:2014:PVN:2623330.2623707}). \\
\textbf{(ii) Real-world networks:} We use seven real-world networks - (1) Zachary's Karate Club (Kar)\footnote{\label{note1}http://www-personal.umich.edu/\texttildelow mejn/netdata/}, (2) Dolphin social network (Dol)\textsuperscript{\ref{note1}}, (3) Les Miserables (Lesmis)\textsuperscript{\ref{note1}}, (4) Books about US Politics (Polbook)\textsuperscript{\ref{note1}}, (5) Word adjacencies (Adjn)\textsuperscript{\ref{note1}}, (6) US Power Grid (Power)\textsuperscript{\ref{note1}} and (7) DBLP collaboration network (Dblp)\footnote{http://snap.stanford.edu/data/}. Table \ref{tab:1} summarises the statistics of the networks.

Note that we do not require the ground-truth community structure since our primary aim is to deceive a community detection algorithm so that after rewiring, the community affiliation of target nodes remains unrevealed.

\begin{table}[!t]
\centering
\caption{Statistics of the real-world networks ($|V|$ and $|E|$ represent the number of nodes and edges, respectively; $\langle k \rangle$ ($k_{max}$) represents the average (maximum) degree of nodes).}
\label{tab:1}
\vspace{-3mm}
\scalebox{1}{
\begin{tabular}{ccccc}
\hline
Network & $\vert V \vert$ &  $\vert E \vert$ & $\langle k \rangle$ &  $k_{max}$ \\
\hline
Kar   & 34  & 78 & 4.59 & 17              \\
Dol   & 62  & 159 & 5.13 & 12               \\
Lesmis & 77 & 154 & 6.60 & 36                  \\
Polbook & 105  & 441 & 8.40 & 25                \\
Adjnoun & 112 & 425 & 7.60 & 49                \\
Power & 4,941 & 6,594 & 2.67 & 19             \\
Dblp & 317,080 & 1,049,866 & 4.93 & 343         \\\hline
\end{tabular}}
\vspace{-5mm}
\end{table}

\vspace{-3mm}
\subsection{Baseline Methods}
We compare NEURAL with four baseline methods:
\begin{enumerate}
    \item {\bf Random algorithm} updates the network by randomly selecting the type of edge update (edge addition/deletion), along with the end nodes.
    \item {\bf Nagaraja} algorithm \cite{inproceedings} updates the network by adding edges between nodes selected on the basis of vertex-centrality measures.
    \item {\bf DICE} \cite{Waniek_2018} updates the network by randomly adding inter-community edges or deleting intra-community edges.
    \item {\bf SADDEN} \cite{8118127} updates the network by maximizing the \textit{safeness gain} in every iteration of edge update based on greedy optimization. 
\end{enumerate}

\vspace{-3mm}
\subsection{Community Detection Algorithms}
We consider six diverse and widely used community detection algorithms: Louvain (Louv) \cite{Blondel_2008}, 
WalkTrap (Walk) \cite{10.1007/11569596_31},
Greedy \cite{PhysRevE.70.066111},
InfoMap (Info) \cite{Rosvall1118},
Label Propagation (Labprop) \cite{Raghavan_2007},
and Leading Eigenvectors (Eig) \cite{PhysRevE.74.036104}. 
Note that none of these algorithms use Permanence as a metric for optimization. Therefore, NEURAL is agnostic to the underlying mechanism of these algorithms.  


\subsection{Evaluation Metrics}
Here, we briefly describe the metrics used to evaluate the community deception methods. $\uparrow$ ({resp.} $\downarrow$) indicates higher ({\em resp.} lower) the value of the metric, better the performance.\\
{\bf (i) Normalized Mutual Information (NMI) $\downarrow$ \cite{Danon_2005}:} To check how much the deception methods are able to hide a particular target community \textit{C} in the network, we calculate the NMI score between the original community structure of the network, $CS = (C_{1}, C_{2}, ...C_{k})$ and the new community structure obtained from a community detection algorithm on the updated network, $CS' = (C'_{1}, C'_{2}, ...C'_{k'})$. The  metric ranges from $0$ (suggesting no overlap between $CS$ and $CS'$) to $1$ (suggesting a complete overlap between $CS$ and $CS'$).\\ 
{\bf (ii) Modified Normalized Mutual Information (MNMI) $\downarrow$:} 
    For large networks, hiding a target community \textit{C} may not have a major effect on the {\em other communities} which are not in immediate contact with $C$. As a result, to capture how effective a deception method is in hiding $C$, we may need to measure NMI between the community memberships of nodes in the target communities and their immediate neighbors before and after the edge updates. We call this metric MNMI. Its range is same as that of NMI.\\
{\bf (iii) Community Splits (CommS) $\uparrow$:}  We propose this metric to define the number of communities in $CS'$ containing the nodes of the target community \textit{C} in the updated network $G'$. It ranges  from 1 (all nodes in $C$ remain in one community in $CS'$) to $\vert CS' \vert$ (all nodes in $C$ get distributed into different communities of $CS'$). The higher the value of CommS, the wider would be the split of the nodes in $C$, thereby increasing the deception of the target community.\\
    \scalebox{0.95}{
        CommS = $\sum_{C'_{i} \in CS'}h(C'_{i}, C)$; $h(C'_{i}, C) = \begin{cases}
          1 & V_{C} \cap V_{C'_{i}} \neq \phi \\
          0 & V_{C} \cap V_{C'_{i}} = \phi
        \end{cases}
    $}
    where $V_{C}$ represents the set of nodes belonging to $C$, and $V_{C_{i}}$ represents set of nodes belonging to community $C_{i}\in CS'$.\\
    {\bf (iv) Community Uniformity (CommU) $\uparrow$:} We propose this metric to capture how nodes in the target community $C$ get distributed among communities in the new community structure $CS'$. It is obtained by calculating the entropy of target community's nodes present among the communities in $CS'$ as follows: 
   $ \text{CommU =} \sum_{C'_{i} \epsilon CS'}-\frac{\vert V_{C, C'_{i}} \vert}{\vert V_{C} \vert} \log\frac{\vert V_{C, C'_{i}} \vert}{\vert V_{C} \vert}$,
    where $\vert V_{C, C'_{i}} \vert$ represents the number of nodes in $C$ present in $C'_{i}\in CS'$, and $\vert V_{C} \vert$ represents the total number of nodes present in $C$.
    It ranges from 0 (when all nodes of $C$ remain in one community of $CS'$) to $\mathcal{\log \vert CS' \vert}$ (when all nodes of $C$ get distributed into different communities of $CS'$). 

\begin{figure}[!t]
  \centering
  \includegraphics[width=\columnwidth]{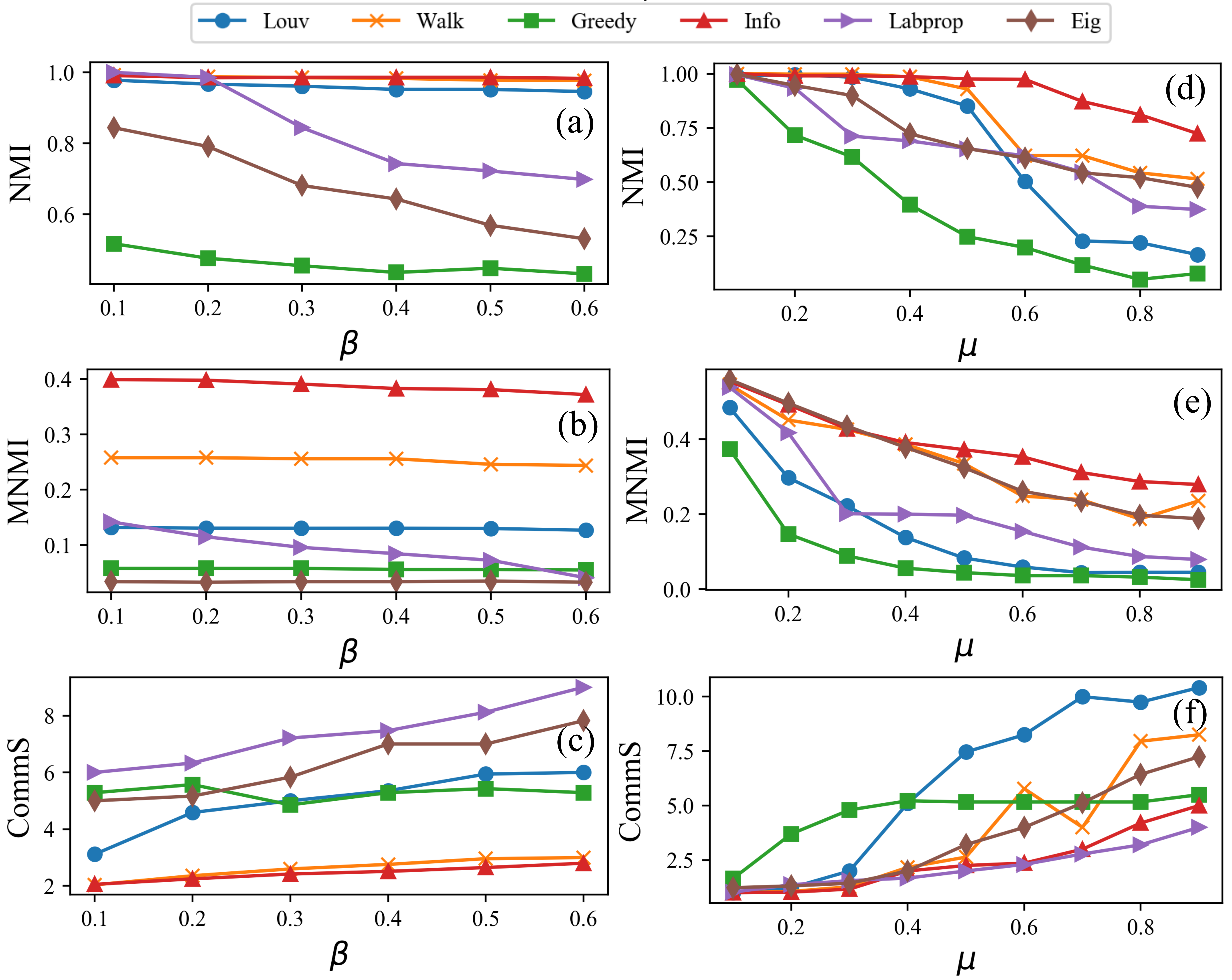}
  \vspace{-5mm}
  \caption{NMI, MNMI and CommS on the default synthetic network by varying $\beta$ in (a)-(c) and $\mu$ in (d)-(f), keeping $\beta=0.3|V_c|$, where $|V_C|$ is the number of nodes in the target community.}
  \label{fig:4}
  \vspace{-3mm}
\end{figure}

\begin{figure*}[!ht]
  \centering
  \includegraphics[width=0.9\textwidth]{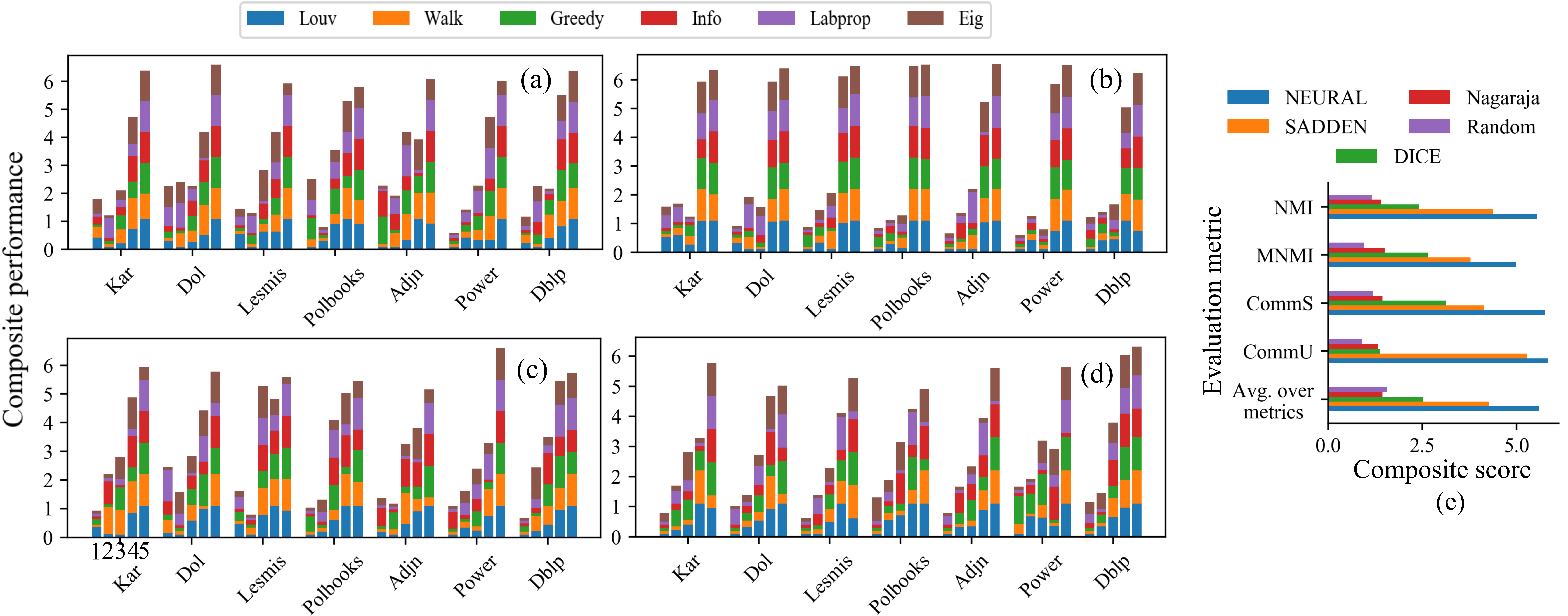}
  \caption{(Color online) Composite performance of the five competing community deception methods based on (a) NMI, (b) MNMI, (c) CommS, and (d) CommU. Bars in each group (under each dataset) are ordered as follows: (1) Random, (2) Nagaraja, (3) DICE,  (4) SADDEN, and (5) NEURAL (as shown in Fig. (c)). Fig.(e) shows the composite performance of each competing method based on every evaluation metric averaged over all the datasets.}
  \label{fig:6}
  \vspace{-5mm}
\end{figure*}

\vspace{-3mm}
\section{Quantitative Evaluation}
Here we present the quantitative analysis of experimental results on both synthetic and real-world networks.

\begin{table}[!t]
 \centering
 \caption{Comparison on the default LFR network, keeping $\beta=0.3|V_{C}|$, where $V_{C}$ is the size the target community.}\label{tab:synthetic}
 \vspace{-1mm}
 \scalebox{1}{
 \begin{tabular}{c|c|c|c|c}
 \hline
      Method & NMI & MNMI  & CommS & CommU  \\\hline
      Random & 0.99 & 0.97 & 1.13 & 0.15 \\  
      Nagaraja & 0.99 & 0.28 & 1.21 & 0.86 \\  
      DICE & {\bf 0.98} & 0.90 & 1.33 & 0.81 \\
      SADDEN & {\bf 0.98} & {\bf 0.26} & 3.64 & 0.73\\
      NEURAL & {\bf 0.98} & {\bf 0.26} & {\bf 3.80} & {\bf 0.94}\\\hline
 \end{tabular}}
 \vspace{-5mm}
\end{table}

\subsection{Evaluation on Synthetic Networks}
We use the default LFR network, set the budget $\beta$ as the fraction of nodes in the target community $C$ and vary the fraction from 0.1 to 0.6. The result is averaged over 20 synthetic networks, 5 randomly selected target communities and 10 runs for each target community. 
Figs. \ref{fig:4}(a)-(c) show that with an increase of $\beta$, NEURAL is able to hide $C$ better (NMI, MNMI scores decrease and CommS scores increase) showing a parallel between the allowed budget and its effect on community deception. 

We further conduct experiments by varying the parameter $\mu$ of LFR network from 0.1 to 0.9. 
Figs. \ref{fig:4}(d)-(f) show that with an increase in $\mu$, the nodes in $C$ are concealed more by NEURAL (NMI, MNMI scores decrease and CommS scores increase). The above observation matches the expectation that it would be easier to hide a target community which has more sparse intra-community connections than the inter-community connections.

Table \ref{tab:synthetic} shows that NEURAL delivers comparable (and sometimes better) accuracy on the default synthetic network.

\subsection{Evaluation on  Real-world Networks}
In case of experiments on real-world networks, we fix $\beta$ to $30\%$ of the  size of the target community $C$ (i.e., $\beta=0.3|V_{C}|$). The results reported here are obtained by averaging the performance considering each of the communities as target community at a time and over 10 runs for each target community.
 
For a compact visualization,  we rank five competing community deception methods as follows: for each evaluation metric and each  community detection algorithm, we normalize their scores (using min-max normalization) so that the best performing method gets score $1$. Now if a competing method outperforms others  by deceiving {\em all} the six community detection algorithms w.r.t. that evaluation metric, it will secure a composite score of $6$. 

Figs. \ref{fig:6}(a)-(d) show the composite performance  across all the evaluation metrics. Fig. \ref{fig:6}(e) shows the composite performance of individual competing methods averaged over all the datasets. We observe that 
NEURAL outperforms others with a significant margin - NEURAL achieves a composite score of $5.53$ (averaged over all the evaluation metrics and datasets), outperforming Random, Nagaraja, DICE and SADDEN by $376.72\%$, $286.71\%$, $128.51\%$, and $26.54\%$, respectively  
(see supplementary for the raw accuracy scores over all the datasets). Note that SADDEN turns out to be highly competitive, sometimes showing marginal improvement over NEURAL. However, in general NEURAL is better than others (Fig. \ref{fig:6}(e)). We also consider hiding individual target nodes instead of communities. Refer supplementary for the same.  

\vspace{-3mm}
\subsection{Non-uniform Budget for Edge Updates}
Till now, we have reported the results with a unified budget $\beta$ for all the edge update operations. In this section, we extend NEURAL with non-uniform budget wherein separate budget constraints are applied to the two types of allowed edge updates (as elaborated in Section \ref{sec:algo}): (i)  $\beta_{D}$ for intra-community edge deletion, and (ii) $\beta_{A}$ for inter-community edge addition. Such an analysis could be useful in situations wherein the costs incurred while deleting an intra-community edge and adding an inter-community edge are different. We perform experiments under two different settings of $\beta_{D}$, $\beta_{A}$: (i) $\beta_{D} = 0.3\beta,$ $\beta_{A} = 0.7\beta$, and (ii) $\beta_{D} = 0.3\beta,$ $\beta_{A} = 0.7\beta$ (we fix $\beta$ as default, i.e., $30\%$ of the  size of the target community $C$). Tables \ref{tab:add_res} and \ref{tab:del_res} provide raw accuracy values (by averaging over all the communities) for NEURAL and SADDEN (the best baseline, extending it in a similar manner) on Karate real-world network (see supplementary for others) for the two settings mentioned above. We observe that NEURAL outperforms SADDEN in most cases. 

\begin{table}[!t]
\centering
\caption{Accuracy of two competing community deception methods: (1) S: SADDEN (best baseline), and (2) N: NEURAL over Karate, such that $\beta_D = 0.3\beta$; $\beta_A = 0.7\beta$}
\vspace{-3mm}
\label{tab:add_res}
\scalebox{0.9}{\begin{tabular}{|c|c|c|c|c|c|c|c|c|}
\hline
Comm. Det. & \multicolumn{2}{c|}{NMI} & \multicolumn{2}{c|}{MNMI} & \multicolumn{2}{c|}{CommS} & \multicolumn{2}{c|}{CommU} \\ \cline{2-9}
Algo. & S & N  & S & N & S & N & S & N\\ \hline
Louv  & 0.94 & {\bf 0.75} & 0.50 & {\bf 0.30} & 1.50 & {\bf 2.50} & 0.27 & {\bf 0.64}\\ \hline
Walk & 0.78 & {\bf 0.74} & 0.37 & {\bf 0.30} & 1.40 & {\bf 2.20} & 0.57 & {\bf 0.96} \\ \hline
Greedy & 0.77 & {\bf 0.64} & 0.35 & {\bf 0.31} & 2.00 & {\bf 2.67} & 0.61 & {\bf 0.74} \\ \hline
Info & 0.83 & {\bf 0.69} & 0.52 & {\bf 0.45} & {\bf 1.33} & 1.00 & 1.75 & {\bf 2.84} \\ \hline
Labprop & 0.84 & {\bf 0.00} & 0.09 & {\bf 0.00} & 1.50 & {\bf 4.00} & {\bf 1.56} & 0.40 \\ \hline
Eig & 0.95 & {\bf 0.82} & 0.41 & {\bf 0.32} & {\bf 1.50} & {\bf 1.50} & 1.53 & {\bf 1.64} \\ \hline
\end{tabular}}
\vspace{-3mm}
\end{table}

\begin{table}[!t]
\centering
\caption{Accuracy of two competing community deception methods: (1) S: SADDEN (best baseline), and (2) N: NEURAL over Karate, such that $\beta_D = 0.7\beta$; $\beta_A = 0.3\beta$}
\vspace{-3mm}
\label{tab:del_res}
\scalebox{0.9}{\begin{tabular}{|c|c|c|c|c|c|c|c|c|}
\hline
Comm. Det. & \multicolumn{2}{c|}{NMI} & \multicolumn{2}{c|}{MNMI} & \multicolumn{2}{c|}{CommS} & \multicolumn{2}{c|}{CommU} \\ \cline{2-9}
Algo. & S & N  & S & N & S & N & S & N\\ \hline
Louv  & 0.94 & {\bf 0.84} & 0.54 & {\bf 0.50} & 1.50 & {\bf 2.00} & 0.79 & {\bf 0.87}\\ \hline
Walk & 0.79 & {\bf 0.78} & 0.49 & {\bf 0.44} & 1.40 & {\bf 2.00} & 0.39 & {\bf 0.86} \\ \hline
Greedy & 0.81 & {\bf 0.66} & 0.34 & {\bf 0.31} & 2.10 & {\bf 2.67} & 0.66 & {\bf 0.74} \\ \hline
Info & {\bf 0.80} & 0.93 & 0.44 & {\bf 0.42} & {\bf 2.00} & 1.00 & 0.85 & {\bf 2.39} \\ \hline
Labprop & 0.68 & {\bf 0.65} & 0.43 & {\bf 0.13} & 1.33 & {\bf 1.67} & {\bf 1.76} & 1.45 \\ \hline
Eig & 0.92 & {\bf 0.89} & 0.45 & {\bf 0.33} & 1.25 & {\bf 1.75} & 0.95 & {\bf 2.33} \\ \hline
\end{tabular}}
\vspace{-3mm}
\end{table}

\vspace{-3mm}
\section{Qualitative Evaluation}
To interpret the rewiring suggested by NEURAL and SADDEN (top two methods), we further take three real-world attributed networks. 
Unless otherwise state, we only consider deletion of edges (as addition of a new edge does not make any sense for these networks).
Louvain algorithm is used for community detection, and the largest community is considered as the target community. 


\vspace{-3mm}
\subsection{Citation Network}
\begin{figure}[!ht]
  \centering
  \includegraphics[width=\columnwidth]{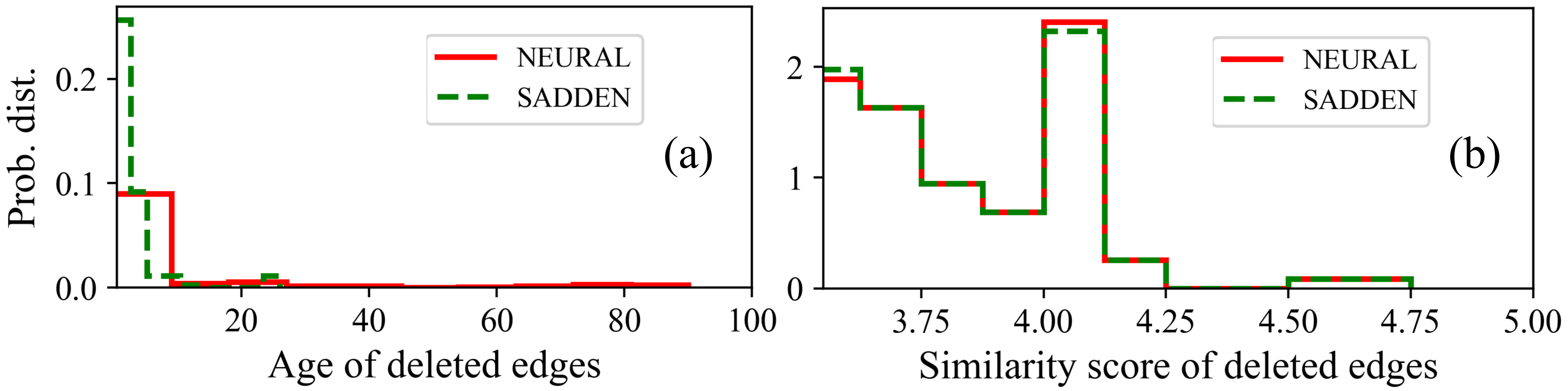}
  \vspace{-5mm}
  \caption{(a) Age and (b) similarity score distribution of edges  selected by NEURAL and SADDEN from Citation and Terrorist networks, respectively.}
  \label{fig:qual}
  \vspace{-5mm}
\end{figure}

We consider $6,320$ papers published in Physical Review Journals as nodes and $10,000$ citation interactions (we ignore directionality) among them as 
edges\footnote{https://journals.aps.org/datasets}. After hiding the target community (largest community) we observe that NEURAL tends to pick up
those citation interactions (or edges) whose {\em age} (defined by the difference between the publication years of citing and cited papers) is relatively high (we believe that these edges have much more importance in terms of keeping the identity of the target community intact, being connected to papers (or nodes) published earlier than most in literature). NEURAL performs better than SADDEN in terms of updating more edges of such kind (Fig. \ref{fig:qual}(a)). 

We further measure the correlation (Spearman's $\rho$ and Kendall's $\tau$) of $138$ edges selected and ranked by NEURAL and those ranked by their age (ground-truth) (similar correlation for $138$ edges returned by SADDEN). Table \ref{tab:rank_corr} shows that NEURAL outperforms SADDEN. Moreover, NEURAL returns the top three edges based on their age present in the target community within top 20 of the rank list, whereas SADDEN is unable to return a single such edge within the 138 edges returned. 

 
    \subsection{Terrorist Network} 
    We use the \textit{Global Terrorism Database}\footnote{https://www.start.umd.edu/gtd/} to create a network of terrorist group associations. This dataset consists of $191,465$ terrorist events around the world between 1970-2018. In order to create the network, we compute the similarity between two terrorist groups based on their activities. To quantify similarity between two groups, we use five attributes: (i) severity of the attack (number of casualties), (ii) attacking strategy used in majority events, (iii) type of weapon used in majority events, (iv) peak year of attacks, and (v) the target type in majority events. Thus, two terrorist groups are associated with a link if the similarity score is greater than or equal to 2.5 (out of 5). This gives rise to a network having $3,616$ nodes as terrorist groups and $22,141$ unweighted edges as association links among these groups. 
    
     After hiding the target community (largest community), we observe that NEURAL first picks up those edges which have higher similarity scores, removing a link between two highly similar terrorist groups. NEURAL performs better compared to SADDEN in terms of providing more edges with high similarity scores (Fig. \ref{fig:qual}(b)). We further measure the rank correlation between $93$ edges returned and ranked by NEURAL with those ranked by their similarity scores (ground-truth) (similarly for $93$ edges returned by SADDEN). Table \ref{tab:rank_corr} shows that NEURAL once again outperforms SADDEN in terms of returning edges whose similarity score is high.

     \subsection{Breast Cancer Network }
     Breast cancer \footnote{This analysis was conducted by two professional biologists.} is considered a leading cause of morbidity and mortality among women worldwide. Above 12\% of the women in the United States are diagnosed with breast cancer during their lifetime \cite{waks2019breast}. Alteration of gene regulation has been widely studied in this context \cite{sengupta2013topological}, with a special focus on dynamic changes in gene co-expression modules. Under the Cancer Genome Atlas (TCGA) program, a community-scale effort has been directed towards multi-omic molecular profiling of breast tumors in hundreds of patients \cite{cancer2012comprehensive}. We use Fragments Per Kilobase of transcript per Million mapped reads (FPKM) normalized gene expression data from TCGA to understand if disguising community affiliation plays a role in the pathogenesis of critical diseases such as cancers. To achieve this, we construct a control and a cancer-specific co-expression network based on transcriptomic profiles of 1,097 normal (as controls) and $113$ tumor samples obtained from the TCGA repository. Both networks spanned the same set of 1,000 genes (1,000 nodes). Two nodes are connected by an edge when the Pearson's correlation coefficient computed across the entire spectrum of control/tumor samples qualifies a cut-off value of $0.6$ ($12,161$ edges). Deleterious mutations in cancer cause wide-spread loss-of-function events, which are often manifested by changes in gene expression levels. 
     
     We employ NEURAL and SADDEN to retrieve co-expressions (edges), whose disappearance fosters community disintegration. NEURAL and SADDEN could pin-point $15$ and $10$ rewirings in the form of edge deletion, respectively, which could be cross-validated w.r.t. the cancer-specific network. Quite strikingly, $5$ out of the $15$ correctly predicted deletions by NEURAL, harbors BMP2 inducible kinase (BPMP2K) as one of the nodes. We find definitive studies implicating this molecule in breast cancer \cite{buraschi2012decorin}.  We fail to find any literature support for the novel gene Z97832.2 that was relatively enriched ($3$ out $10$ rewirings) among SADDEN predicted rewirings. We also measure how accurate NEURAL and SADDEN are to predict the ground-truth edges w.r.t the cancer-specific network. Table \ref{tab:rank_corr} shows that NEURAL outperforms SADDEN on four evaluation measures. 
     To this end, we conclude that NEURAL-led investigation of genome-scale molecular networks holds significant promise in understanding genetic diseases such as cancers.

     
\begin{table}[!t]
    \centering
    \caption{Rank correlation for citation and terrorist networks, and accuracy for breast cancer network.}
    \scalebox{0.9}{
    \begin{tabular}{|c|c|c|c|c|c}
    \cline{1-5}
         \multirow{2}{*}{Method} & \multicolumn{2}{c|}{Citation} & 
         \multicolumn{2}{c|}{Terrorist}  \\\cline{2-5}
         &\multicolumn{2}{c|}{Spearman's $\rho$}& \multicolumn{2}{c|}{Kendall's $\tau$} &   \\\cline{1-5}
         SADDEN  & 0.12 & 0.21 & 0.00 & 0.07 & Rank correlations are \\\cline{1-5}
        NEURAL & {\bf 0.16} & {\bf 0.41} & {\bf 0.06} & {\bf 0.29} & statistically significant  \\\cline{1-5}\cline{1-5}
        
        \multirow{2}{*}{ } & \multicolumn{4}{c|}{Breast cancer} & with $p$-value$>0.8$ \\\cline{2-5}
        & MAP & F1 score  & nDCG & AUC & \\\cline{1-5}
        SADDEN & 0.004 & 0.20 & 0.47 & 0.30 \\\cline{1-5}
        NEURAL & {\bf 0.006} & {\bf 0.29} & {\bf 0.54} & {\bf 0.39} \\\cline{1-5}
    \end{tabular}}
    \label{tab:rank_corr}
    \vspace{-3mm}
\end{table}

\section{Conclusion}
This paper addressed the problem of {\em community deception} -- outwitting  community detection algorithms from discovering the community affiliation of nodes in a target community. Our major contributions are as follows: (i) we {\bf formalized} the problem and called it {\em Hide and Seek Community} (HSC); (ii) we proposed a {\bf novel objective function} (Permanence loss) which has been analyzed theoretically; (iii) we proposed NEURAL, a {\bf novel greedy strategy} to optimize Permanence loss; (iv) NEURAL turned out to be {\bf more efficient} than the  baselines; and (v) NEURAL unfolded different {\bf meta-information of edges} which would otherwise not have been possible to explain just by analyzing the network structure. In particular, NEURAL showed promise in the analysis of genome-scale molecular networks.
 
\section*{Acknowledgement}
The work was partially supported by the Ramanujan Fellowship and DST (ECR/2017/00l691). T. Chakraborty would like to acknowledge the support of CAI, IIIT-Delhi.

\bibliographystyle{IEEEtran}
\bibliography{IEEEabrv,references_TETC} 

\vspace{-10mm}
\begin{IEEEbiography}[{\includegraphics[width=1in,height=1.25in,clip,keepaspectratio]{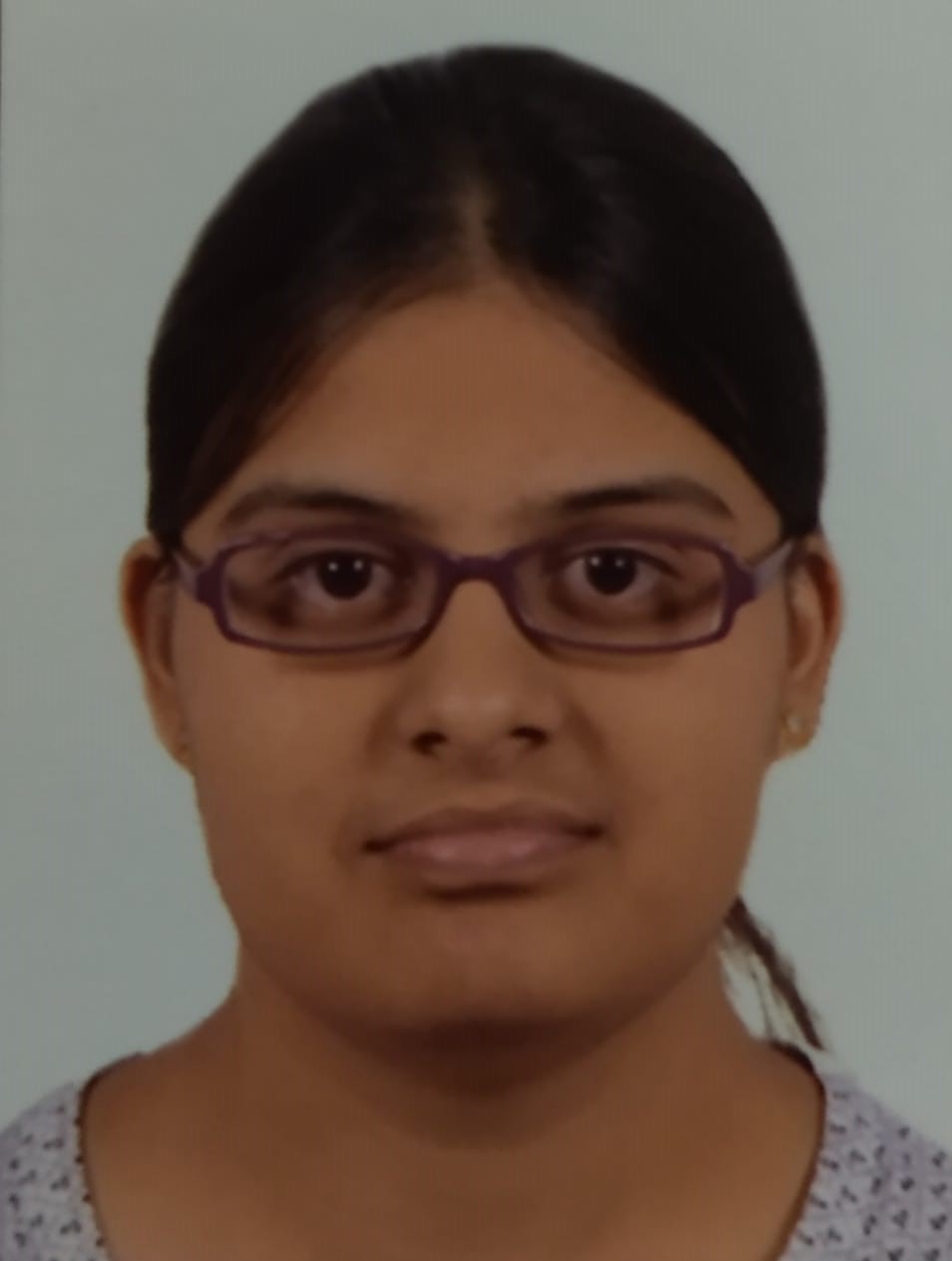}}]{Shravika Mittal} is a senior undergraduate student in Computer Science and Engineering at IIIT-Delhi. Her research interests include Social Network Analysis, Network Science, and Natural Language Processing. She has received the Dean's list for Excellence in Academics, and Innovation in Research and Development. 
\end{IEEEbiography}

\vspace{-10mm}
\begin{IEEEbiography}[{\includegraphics[width=1in,height=1.25in,clip,keepaspectratio]{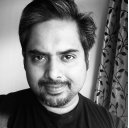}}]{Debarka Sengupta} 
received his Ph.D. from Jadavpur University. 
Before joining
IIIT-D, he worked as an INSPIRE Faculty at Indian Statistical Institute. 
He consulted and advised a number of technology and service-based firms including IPsoft, Datanomers, CoreCompete and Applied Research Works on various data science and business analytics projects. 
He has twice been nominated for the prestigious INSPIRE Faculty award - in 2014 and 2016. 

\end{IEEEbiography}

\vspace{-15mm}
\begin{IEEEbiography}[{\includegraphics[width=1in,height=1.25in,clip,keepaspectratio]{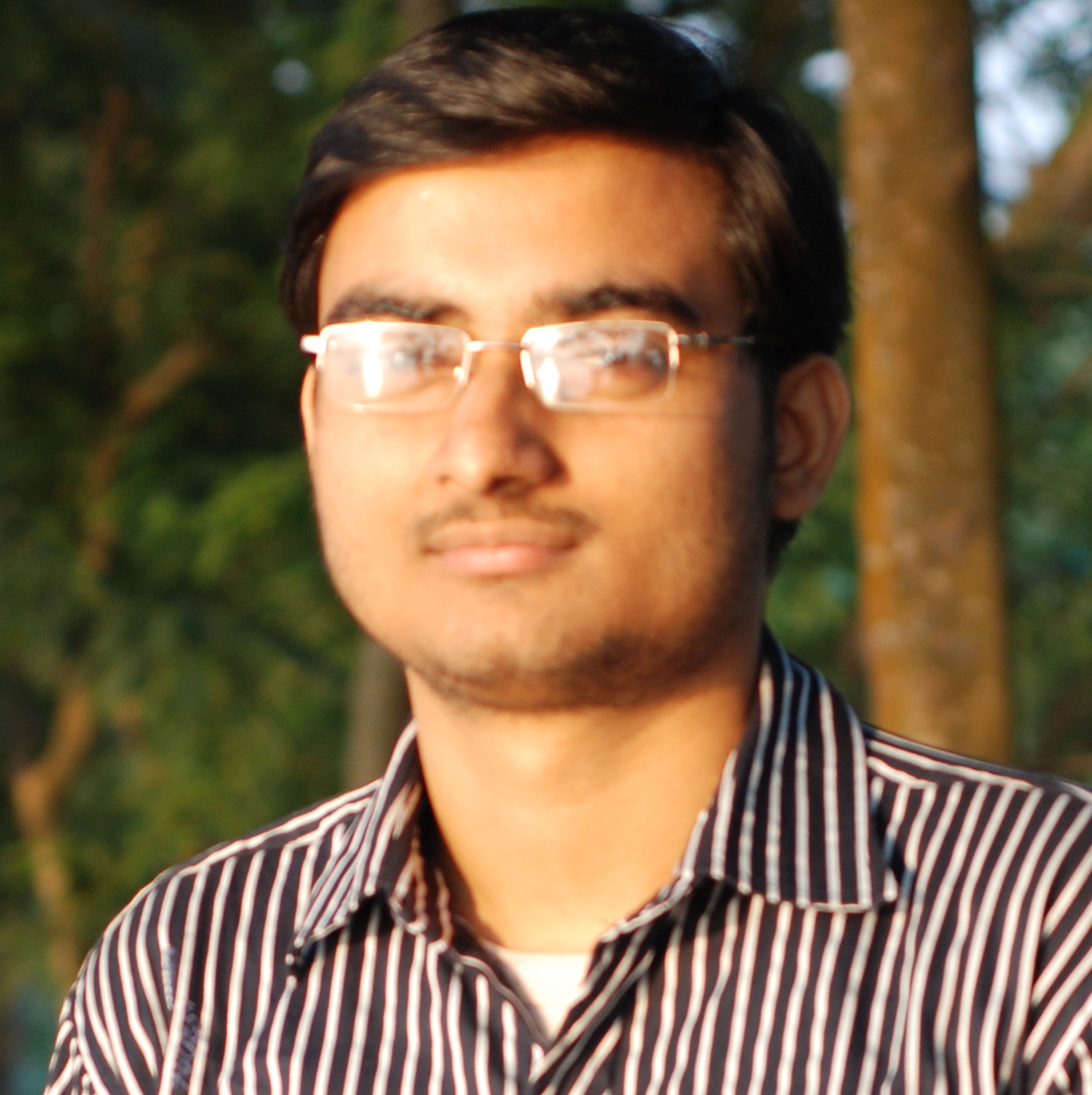}}]{Tanmoy Chakraborty} is an Assistant Professor and a Ramanujan Fellow at the Dept. of Computer Science and Engineering, IIIT-Delhi, India, where he leads a research group, called LCS2 (\url{http://lcs2.iiitd.edu.in/}). His primary research interests include Social Network Analysis, Data Mining, and Natural Language Processing. He has received several awards including Google Indian Faculty Award, Early Career Research Award, DAAD Faculty award. More details at \url{http://faculty.iiitd.ac.in/~tanmoy/}.
\end{IEEEbiography}


\newpage
\thispagestyle{plain}
\makeatletter
\twocolumn[\centering
\vspace{.6ex}
\LARGE Hide and Seek: Outwitting Community Detection Algorithms \\ (Supplementary Materials) \\
\large Shravika Mittal, Debarka Sengupta, Tanmoy Chakraborty \\
\{shravika16093, debarka, tanmoy\}@iiitd.ac.in \par \bigskip
\makeatother 
]

\section{Permanence loss is Submodular and Monotone}\label{sec:proof}
In this section, we prove that the proposed objective function is submodular and monotone w.r.t. the number of edge updates.

{\bf 
\begin{theorem}\label{theorem:sub}
    Permanence loss $\mathcal{P}_{l}$ is submodular w.r.t. the addition of an inter-community edge.
\end{theorem}}
\begin{IEEEproof}
    Let $A$ be the set of inter-community edges that are being considered to be added into the network $G$ using Theorem \ref{theorem:3}. Let $x_{1}$ and $x_{2}$ be two other such inter-community edges where $x_{1}, x_{2} \notin A$. 
    $\mathcal{P}_{l}(A \cup \{x_{1}\})$ ({\em resp.} $\mathcal{P}_{l}(A \cup \{x_{2}\})$) represents the permanence loss due to the addition of  the inter-community edge set $A \cup \{x_{1}\}$ ({\em resp.} $A \cup \{x_{2}\}$) (given by \eqref{eq:5}). Therefore,
    \begin{equation*}\small
        \begin{split}
            \mathcal{P}_{l}(A \cup \{x_{1}\}) + \mathcal{P}_{l}(A \cup \{x_{2}\}) &= 2\sum_{i \epsilon A}\mathcal{P}_{l}(i) + \mathcal{P}_{l}(x_{1}) + \mathcal{P}_{l}(x_{2})\\
    &= \mathcal{P}_{l}(A \cup \{x_{1}, x_{2}\}) + \mathcal{P}_{l}(A)
        \end{split}
    \end{equation*}
     
    This shows that Permanence loss is submodular w.r.t. the addition of an inter-community edge. 
\end{IEEEproof}
{\bf 
\begin{theorem}\label{theorem:sub_1}
    Permanence loss $\mathcal{P}_{l}$ is submodular w.r.t. the deletion of an intra-community edge.
\end{theorem}}
\begin{IEEEproof}
    Let $A$ be the set of intra-community edges that are being considered to be added into the network $G$ using Theorem \ref{theorem:2}. Let $x_{1}$ and $x_{2}$ be two other such intra-community edges such that $x_{1}, x_{2} \notin A$. 
    $\mathcal{P}_{l}(A \cup \{x_{1}\})$ ({\em resp. $\mathcal{P}_{l}(A \cup \{x_{2}\})$)}  represents the permanence loss due to the deletion of the set of intra-community edges $A \cup \{x_{1}\}$ ({\em resp.} $A \cup \{x_{2}\}$) (given by \eqref{eq:del}). Therefore,
    \begin{equation*}\small
        \begin{split}
            \mathcal{P}_{l}(A \cup \{x_{1}\}) + \mathcal{P}_{l}(A \cup \{x_{2}\}) &= 2\sum_{i \epsilon A}\mathcal{P}_{l}(i) + \mathcal{P}_{l}(x_{1}) + \mathcal{P}_{l}(x_{2})\\
    &= \mathcal{P}_{l}(A \cup \{x_{1}, x_{2}\}) + \mathcal{P}_{l}(A)
    \end{split}
    \end{equation*}
    This shows that Permanence loss is submodular w.r.t. the deletion of an intra-community edge. 
\end{IEEEproof}
{\bf 
\begin{theorem}\label{theorem:mon}
    Permanence loss $\mathcal{P}_{l}$ is monotone w.r.t. the addition of an inter-community edge.
\end{theorem}}
\begin{IEEEproof}
    Let $A$ be the set of inter-community edges obtained using Theorem \ref{theorem:3}. Let $B$ be the set of inter-community edges such that $B = A \cup \{x_{1}\}$, where $x_{1}$ is another inter-community edge considered to update the network for community deception. Since
    $A \subseteq B$, 
    $\mathcal{P}_{l}(B) = \mathcal{P}_{l}(A) + \mathcal{P}_{l}(x_{1}) \geq \mathcal{P}_{l}(A)$
    using \eqref{eq:5}, which proves that the Permanence loss is monotone w.r.t. to the addition of an inter-community edge.
\end{IEEEproof}
{\bf 
\begin{theorem}\label{theorem:mon_1}
    Permanence loss $\mathcal{P}_{l}$ is monotone w.r.t. the deletion of an intra-community edge.
\end{theorem}}
\begin{IEEEproof}
    Let $A$ be the set of intra-community edges obtained using Theorem \ref{theorem:2}. Let $B$ be the set of intra-community edges such that $B = A \cup \{x_{1}\}$, where $x_{1}$ is another intra-community edge considered to update the network for community deception. Since $A \subseteq B$,
    $\mathcal{P}_{l}(B) = \mathcal{P}_{l}(A) + \mathcal{P}_{l}(x_{1}) \geq \mathcal{P}_{l}(A)$ 
    using \eqref{eq:del}, which proves that the Permanence loss is monotone w.r.t. the deletion of an intra-community edge.
\end{IEEEproof}

\begin{table}[]
    \centering
    \caption{Scores of two methods: (1) SADDEN, and (2) NEURAL for hiding individual target nodes, averaged across 20 different runs.}
    \scalebox{0.9}{\begin{tabular}{c|c|c} \hline
         Network & Score (SADDEN) & Score (NEURAL) \\ \hline
         Kar & 0.62 & \bf{0.69}\\
         Dol & 0.59 & \bf{0.71}\\
         Lesmis & 0.64 & \bf{0.70}\\
         Polbook & 0.71 & \bf{0.75}\\
         Adjnoun & 0.68 & \bf{0.79}\\
         Power & 0.54 & \bf{0.68}\\
         Dblp & 0.64 & \bf{0.71}\\ \hline
    \end{tabular}}
    \label{tab:ind_nodes}
\end{table}

\vspace{-3mm}
\section{Hiding Nodes rather than Communities}
In this section, we address a modified version of our problem statement to hide individual target nodes instead of communities. To hide nodes, we employ our proposed  NEURAL algorithm by treating nodes as singleton communities. Since deleting an intra-community edge would be redundant in case of a singleton community, NEURAL only focuses on adding inter-community edges to hide target nodes. We evaluate our methodology by selecting $0.3|V|$ nodes randomly as targets for 7 real-world networks. Louvain algorithm is used to extract community assignment. Table \ref{tab:ind_nodes} summarises the probability scores of hiding the target nodes for NEURAL and SADDEN (the best baseline). We observe that NEURAL outperforms SADDEN by assigning a different community label for more target nodes.

\begin{figure}[!t]
  \centering
  \includegraphics[scale=0.5]{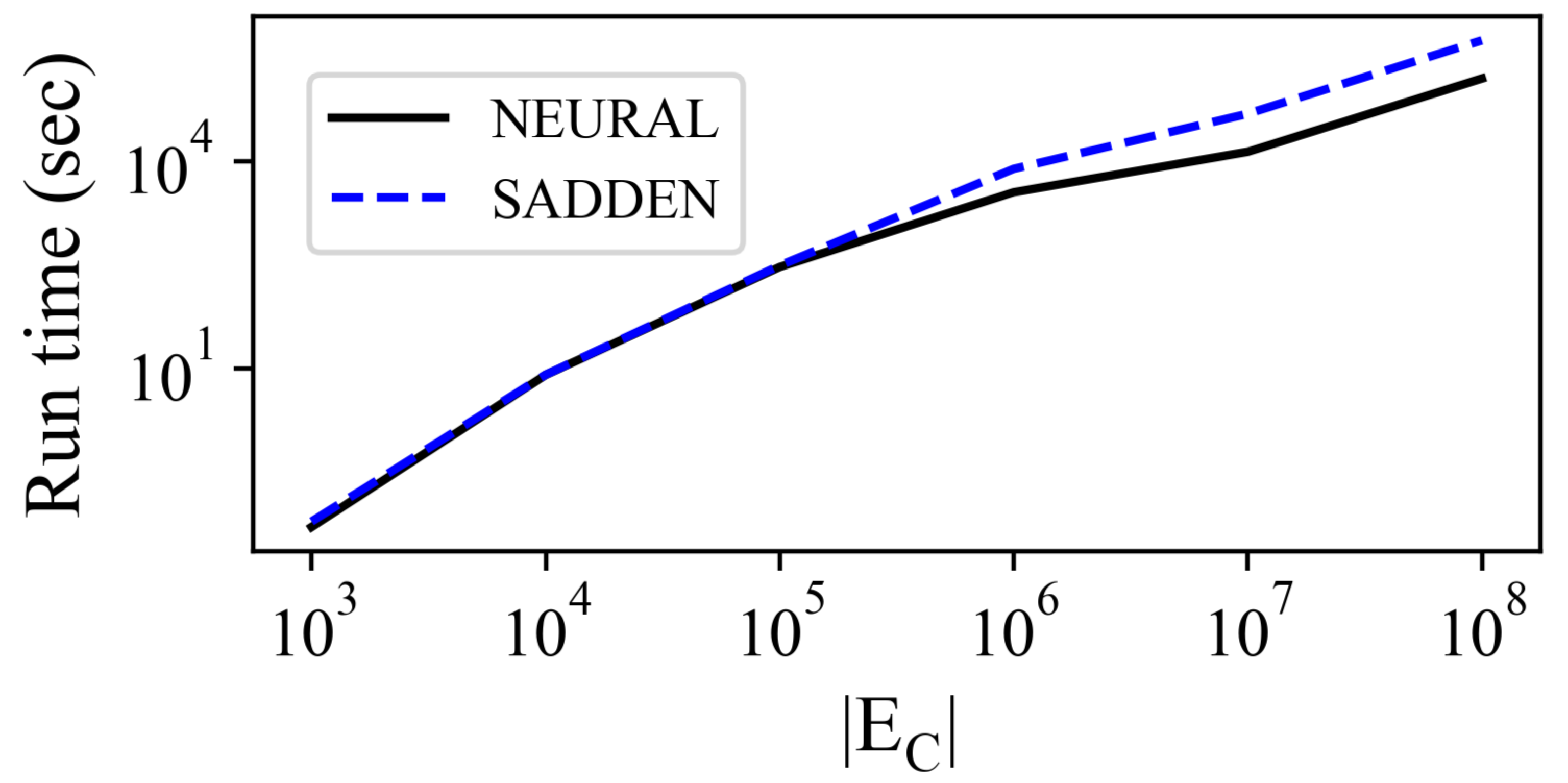}
  \vspace{-5mm}
  \caption{Scalability analysis of NEURAL and SADDEN.}
  \label{fig:scalability}
  \vspace{-5mm}
\end{figure}

\vspace{-5mm}
\section{Scalability Analysis}\label{sec:scale}  
The discussion in Section \ref{sec:algo} has established that the time complexity of NEURAL is $\mathcal{O}(\vert V_{C} \vert + \vert E_{C} \vert)$. To show this empirically, we use the LFR Benchmark \cite{Lancichinetti2008Benchmark} for generating synthetic networks with $\mu=0.4$. Louvain algorithm is used for community detection, and the largest community is considered as the target community such that $\vert E_{C} \vert$ lies within the range $10^{3} - 10^{8}$. The results are averaged over 10 such synthetic networks. We record the run times for two deception strategies (based on greedy optimization): (i) SADDEN and (ii) NEURAL (we fix $\beta$ as default i.e., $\beta=0.3|V_{C}|$). Fig. \ref{fig:scalability} shows that the run time for NEURAL increases linearly with $\vert E_{C} \vert$ thereby verifying the analytical time complexity shown in Section \ref{sec:algo} ($\vert E_{C} \vert >> \vert V_{C} \vert$; $\mathcal{O}(\vert V_{C} \vert + \vert E_{C} \vert)$ $\approx$  $\mathcal{O}(\vert E_{C} \vert)$). We also observe that with an increase in $\vert E_{C} \vert$, NEURAL outperforms SADDEN in terms of its run time.

\begin{table*}[!htb]
\centering
\caption{Accuracy of the five competing community deception methods: (1) Nag: Nagaraja, (2) R: Random, (3) D: DICE, (4) S: SADDEN, and (5) N: NEURAL over 7 real world networks: (A) Karate, (B) Dolphin, (C) Lesmis, (D) Polbooks, (E) Adjn, (F) Power and (G) Dblp.}
\label{tab:realres}
\scalebox{0.8}{\begin{tabular}{|c|c|c|c|c|c|c|c|c|c|c|c|c|c|c|c|c|c|c|c|c|}

\multicolumn{21}{c}{{\bf (A) Karate}}\\\hline
Comm. Det. & \multicolumn{5}{c|}{NMI} & \multicolumn{5}{c|}{MNMI} & \multicolumn{5}{c|}{CommS} & \multicolumn{5}{c|}{CommU} \\ \cline{2-21}
Algo. & Nag & R & D & S & N  & Nag & R & D & S & N & Nag & R & D & S & N & Nag & R & D & S & N\\ \hline
Louv  & 0.94 & 0.88 & 0.92 & 0.82 & {\bf 0.75} & 0.33 & 0.65 & 0.95 & 0.27 & {\bf 0.26} & 1.27 & 1.50 & 1.25 & 2.00 & {\bf 2.25}  & 0.55 & 0.61 & 0.72 & {\bf 0.98} & 0.93 \\ \hline
Walk & 0.82 & 0.78 & 0.76 & {\bf 0.67} & 0.70 & 0.46 & 0.70 & 0.61 & {\bf 0.17} & 0.27 & 1.33 & 1.00 & 1.30 & 1.40 & {\bf 1.41} & 0.82 & 1.52 & 1.54 & {\bf 1.81} & 1.61 \\ \hline
Greedy & 0.82 & 0.82 & 0.78 & 0.77 & {\bf 0.72} & 0.25 & 0.85 & 0.73 & 0.25 & {\bf 0.24} & 1.58 & 1.67 & 2.33 & 2.00 & {\bf 2.67} & 0.53 & 0.43 & 0.59 & 0.58 & {\bf 0.71} \\ \hline
Info & 0.89 & 0.85 & 0.85 & 0.72 & {\bf 0.68} & 0.47 & 0.81 & 0.63 & 0.46 & {\bf 0.17} & 1.23 & 1.00 & 1.00 & 1.33  & {\bf 1.34} & 1.87 & 1.84 & 1.91 & 1.87 & {\bf 2.25} \\ \hline
Labprop & 0.29 & 0.76 & 0.64 & 0.52 & {\bf 0.00} & 0.33 & 0.35 & 0.54 & 0.10 & {\bf 0.00} & 1.49 & 1.33 & 1.75 & 2.00  & {\bf 6.00}     & 0.97 & 0.57 & 0.64 & 0.41  & {\bf 1.31} \\ \hline
Eig & 0.94 & 0.89 & 0.91 & 0.83 & {\bf 0.82} & 0.38 & 0.64 & 0.73 & {\bf 0.30} & 0.33 & 1.02 & 1.00 & 1.50 & {\bf 1.75}  & 1.25 & 2.12 & 0.98 & 1.96 & 1.07 & {\bf 2.14} \\ \hline

\multicolumn{21}{c}{{\bf (B) Dolphin}} \\\hline
Comm. Det. & \multicolumn{5}{c|}{NMI} & \multicolumn{5}{c|}{MNMI} & \multicolumn{5}{c|}{CommS} & \multicolumn{5}{c|}{CommU} \\ \cline{2-21}
Algo. & Nag & R & D & S & N  & Nag & R & D & S & N & Nag & R & D & S & N & Nag & R & D & S & N\\ \hline
Louv & 0.89 & 0.84 & 0.85 & 0.79 & {\bf 0.64} & 0.21 & 0.65 & 0.78 & 0.20 & {\bf 0.18} & 1.31 & 1.40 & 2.00 & 2.60 & {\bf 2.75} & 0.64 & 0.41 & 0.56 & 0.69 & {\bf 0.75} \\ \hline
Walk & 0.77 & 0.83 & 0.78 & {\bf 0.67} & {\bf 0.67} & 0.35 & 0.61 & 0.65 & 0.37 & {\bf 0.24} & 2.04 & 2.75 & 3.00 & 2.00 & {\bf 4.25} & 0.88 & 0.79 & 0.98 & {\bf 1.77} & 1.01 \\ \hline
Greedy & 0.91 & 0.89 & 0.85 & 0.81 & {\bf 0.77} & 0.31 & 0.55 & 0.54 & {\bf 0.20} & 0.27 & 1.36 & 1.50 & 2.00 & {\bf 2.75} & 2.50 & 0.78 & 0.54 & 0.76 & 0.67 & {\bf 1.04} \\ \hline
Info & 0.92 & 0.91 & 0.87 & 0.85 & {\bf 0.82} & 0.27 & 0.66 & 0.59 & 0.31 & {\bf 0.25} & 1.29 & 2.20 & 2.20 & 2.16  & {\bf 3.80} & 0.89 & 0.63 & 0.81 & {\bf 1.34} & 0.87 \\ \hline
Labprop & 0.69 & 0.71 & 0.73 & 0.76 & {\bf 0.67} & 0.11 & 0.74 & 0.36 & 0.12 & {\bf 0.07} & 1.69 & {\bf 2.50} & 1.33 & 2.25 & 1.75 & 1.60 & 1.10 & 1.25 & 0.41 & {\bf 2.00} \\ \hline
Eig & 0.84 & 0.84 & 0.88 & 0.83 & {\bf 0.82} & 0.21 & 0.86 & 0.73 & 0.23 & {\bf 0.16} & 1.65 & 1.00 & 1.50 & 1.80 & {\bf 2.00} & 0.66 & 0.64 & 0.87 & {\bf 1.48} & 1.03 \\ \hline

\multicolumn{21}{c}{{\bf (C) Lesmis}} \\\hline
Comm. Det. & \multicolumn{5}{c|}{NMI} & \multicolumn{5}{c|}{MNMI} & \multicolumn{5}{c|}{CommS} & \multicolumn{5}{c|}{CommU} \\ \cline{2-21}
Algo. & Nag & R & D & S & N  & Nag & R & D & S & N & Nag & R & D & S & N & Nag & R & D & S & N\\ \hline
Louv & 0.96 & 0.90 & 0.89 & 0.89 & {\bf 0.83} & 0.32 & 0.85 & 0.84 & 0.33 & {\bf 0.29} & 1.30 & 1.67 & 2.00 & {\bf 2.33} & 2.16  & 0.69 & 0.54 & 0.88 & {\bf 1.41} & 0.98 \\ \hline
Walk & 0.95 & 0.95 & 0.94 & 0.92 & {\bf 0.89} & 0.55 & 0.94 & 0.65 & 0.43 & {\bf 0.39} & 1.30 & 1.50 & 1.75 & 1.88 & {\bf 2.00} & 1.33 & 0.41 & 0.87 & 1.30 & {\bf 1.77} \\ \hline
Greedy & 0.90 & 0.92 & 0.91 & 0.87 & {\bf 0.82} & 0.37 & 0.65 & 0.78 & 0.33 & {\bf 0.32} & 1.30 & 1.50 & 1.75 & 2.00 & {\bf 2.20}  & 0.77 & 0.66 & 0.78 & 0.85 & {\bf 0.98} \\ \hline
Info & 0.98 & 0.98 & 0.95 & 0.94 & {\bf 0.91} & 0.44 & 0.86 & 0.71 & 0.35 & {\bf 0.29} & 1.17 & 1.11 & 2.13 & 2.00  & {\bf 2.38} & 0.81 & 0.89 & 1.02 & 1.05 & {\bf 1.55} \\ \hline
Labprop & 0.69 & 0.74 & 0.78 & 0.69 & {\bf 0.60} & 0.36 & 0.68 & 0.53 & 0.28 & {\bf 0.16} & 1.47 & 1.86 & 2.50 & 2.00  & {\bf 2.67} & 0.14 & 0.44 & 0.41 & {\bf 1.37} & 0.55 \\ \hline
Eig & 0.97 & 0.96 & {\bf 0.91} & {\bf 0.91} & 0.95 & 0.95 & 0.94 & 0.73 & { \bf 0.34} & 0.41 & 0.96 & 1.17 & {\bf 2.70} & 1.75  & 1.25  & 1.49 & 0.45 & 0.97 & 0.49 & {\bf 1.82} \\ \hline

\multicolumn{21}{c}{{\bf (D) Polbooks}} \\\hline
Comm. Det. & \multicolumn{5}{c|}{NMI} & \multicolumn{5}{c|}{MNMI} & \multicolumn{5}{c|}{CommS} & \multicolumn{5}{c|}{CommU} \\ \cline{2-21}
Algo. & Nag & R & D & S & N  & Nag & R & D & S & N & Nag & R & D & S & N & Nag & R & D & S & N\\ \hline
Louv & 0.98 & 0.99 & 0.95 & {\bf 0.94} & 0.95 & 0.24 & 0.79 & 0.76 & 0.22 & {\bf 0.23} & 1.30 & 1.25 & 1.50 & 1.74 & {\bf 1.75}  & 0.44 & 0.43 & 0.62 & 0.74 & {\bf 0.75} \\ \hline
Walk & 0.97 & 0.95 & 0.94 & {\bf 0.85} & 0.94 & 0.94 & 0.92 & 0.76 & 0.32 & {\bf 0.31} & 1.30 & 1.25 & 1.64 & {\bf 2.75} & 1.75 & 0.88 & 0.79 & 0.82 & 0.96 & {\bf 1.26} \\ \hline
Greedy  & 0.97 & 0.93 & 0.92 & 0.95 & {\bf 0.91} & 0.73 & 0.70 & 0.83 & {\bf 0.28} & 0.31 & 1.12 & 1.50 & 1.80 & 1.25 & {\bf 2.00}  & 0.23 & 0.34 & 0.63 & {\bf 2.56} & 0.95 \\ \hline
Info & 0.99 & 0.99 & 0.98 & 0.96 & {\bf 0.95} & 0.34 & 0.74 & 0.74 & {\bf 0.32} & 0.33 & 1.02 & 1.00 & 2.13 & {\bf 2.33}  & 1.83 & 0.54 & 0.39 & 0.98 & 0.58 & {\bf 1.04} \\ \hline
Labprop & 0.87 & 0.81 & 0.80 & 0.78 & {\bf 0.73} & 0.21 & 0.67 & 0.59 & 0.23 & {\bf 0.17} & 1.50 & 1.00 & 3.00 & 1.67  & {\bf 3.33} & 0.47 & 0.48 & 0.43 & {\bf 2.88} & 0.52 \\ \hline
Eig & 0.94 & 0.92 & 0.93 & {\bf 0.91} & 0.92 & 0.26 & 0.74 & 0.63 & 0.19 & {\bf 0.18} & 1.57 & 1.25 & 1.50 & {\bf 2.25}  & 1.75  & 0.61 & 0.66 & 0.76 & 0.26 & {\bf 0.84} \\ \hline

\multicolumn{21}{c}{{\bf (E) Adjn}} \\\hline
Comm. Det. & \multicolumn{5}{c|}{NMI} & \multicolumn{5}{c|}{MNMI} & \multicolumn{5}{c|}{CommS} & \multicolumn{5}{c|}{CommU} \\ \cline{2-21}
Algo. & Nag & R & D & S & N  & Nag & R & D & S & N & Nag & R & D & S & N & Nag & R & D & S & N\\ \hline
Louv & 0.74 & 0.74 & 0.71 & {\bf 0.62} & 0.64 & 0.23 & 0.73 & 0.72 & 0.25 & {\bf 0.22} & 2.73 & 2.86 & 3.23 & 3.86 & {\bf 4.14}  & 0.89 & 0.77 & 0.85 & 1.03 & {\bf 1.10} \\ \hline
Walk & 0.97 & 0.99 & 0.95 & 0.95 & {\bf 0.94} & 0.73 & 0.85 & 0.79 & 0.73 & {\bf 0.69} & 1.22 & 1.16 & {\bf 2.00} & 1.44 & 1.32 & 0.60 & 0.41 & 0.45 & 0.64 & {\bf 0.83} \\ \hline
Greedy & 0.75 & 0.60 & 0.62 & 0.66 & {\bf 0.58} & 0.41 & 0.69 & 0.64 & 0.22 & {\bf 0.21} & 1.00 & 2.50 & 3.50 & 3.14 & {\bf 3.86}  & 0.89 & 0.68 & 0.97 & 1.10 & {\bf 1.37} \\ \hline
Info & 0.50 & 0.45 & 0.51 & 0.57 & {\bf 0.42} & 0.24 & 0.73 & 0.65 & {\bf 0.02} & 0.05 & 1.75 & 1.00 & 1.16 & 1.50  & {\bf 2.50}   & 0.92 & 0.62 & 0.89 & 0.78 & {\bf 1.08} \\ \hline
Labprop & 1.00 & 1.00 & 0.00 & 1.00 & {\bf 0.00} & 1.00 & 1.00 & 0.00 & 1.00 & {\bf 0.00} & 1.00 & 1.00 & 1.00 & 1.00  & {\bf 6.00} & 4.71 & 1.10 & 1.23 & {\bf 4.72} & 0.25 \\ \hline
Eig & 0.74 & 0.85 & 0.77 & {\bf 0.64} & 0.71 & 0.42 & 0.68 & 0.70 & 0.40 & {\bf 0.38} & 1.04 & 1.30 & 1.60 & {\bf 2.80}  & 1.70  & 1.03 & 0.66 & 0.79 & 0.69 & {\bf 1.42} \\ \hline

\multicolumn{21}{c}{{\bf (F) Power}} \\\hline
Comm. Det. & \multicolumn{5}{c|}{NMI} & \multicolumn{5}{c|}{MNMI} & \multicolumn{5}{c|}{CommS} & \multicolumn{5}{c|}{CommU} \\ \cline{2-21}
Algo. & Nag & R & D & S & N  & Nag & R & D & S & N & Nag & R & D & S & N & Nag & R & D & S & N\\ \hline
Louv & 0.97 & 0.98 & 0.99 & 0.98 & {\bf 0.95} & 0.62 & 0.74 & 0.74 & 0.29 & {\bf 0.05} & 2.10 & 1.10 & 1.50 & 3.00 & {\bf 4.00} & 0.14 & 0.11 & 0.15 & 0.13 & {\bf 0.18} \\ \hline
Walk & 0.98 & 0.99 & 0.97 & 0.93 & {\bf 0.91} & 0.78 & 0.83 & 0.81 & {\bf 0.38} & 0.39 & 4.00 & 2.00 & 3.00 & 20.00 & {\bf 24.00} & 0.34 & 0.48 & 0.51 & 0.41 & {\bf 0.75} \\ \hline
Greedy  & 0.97 & 0.98 & 0.96 & 0.94 & {\bf 0.93} & 0.86 & 0.83 & 0.81 & {\bf 0.03} & 0.07 & 2.50 & 2.50 & 4.00 & 3.00 & {\bf 6.00}  & 0.63 & 0.98 & 1.10 & 0.12 & {\bf 1.15} \\ \hline
Info & 0.98 & 0.98 & 0.98 & 0.97 & {\bf 0.95} & 0.89 & 0.92 & 0.91 & 0.22 & {\bf 0.10} & 1.94 & 2.50 & 2.00 & 1.00  & {\bf 4.00} & 0.74 & 0.55 & 0.61 & {\bf 3.89} & 0.68 \\ \hline
Labprop & 0.94 & 0.97 & 0.92 & 0.91 & {\bf 0.90} & 0.84 & 0.87 & 0.86 & 0.38 & {\bf 0.27} & 4.00 & 3.00 & 5.00 & 7.00 & {\bf 8.00} & 1.05 & 1.05 & 1.14 & 1.24 & {\bf 1.74} \\ \hline
Eig & 0.95 & 0.95 & 0.93 & {\bf 0.76} & 0.87 & 0.54 & 0.54 & 0.48 & 0.08 & {\bf 0.04} & 3.22 & 1.50 & 4.00 & 3.00  & {\bf 7.00}  & 0.45 & 0.45 & 0.62 & 0.66 & {\bf 0.72} \\ \hline

\multicolumn{21}{c}{{\bf (G) Dblp}} \\\hline
Comm. Det. & \multicolumn{5}{c|}{NMI} & \multicolumn{5}{c|}{MNMI} & \multicolumn{5}{c|}{CommS} & \multicolumn{5}{c|}{CommU} \\ \cline{2-21}
Algo. & Nag & R & D & S & N  & Nag & R & D & S & N & Nag & R & D & S & N & Nag & R & D & S & N\\ \hline
Louv & 1.00 & 0.99 & 0.99 & 0.98 & {\bf 0.97} & 0.39 & 0.42 & 0.35 & {\bf 0.28} & 0.31 & 3.75 & 3.50 & 4.25 & 5.00 & {\bf 5.33} & 0.08 & 0.05 & 0.10 & 0.15 & {\bf 0.16} \\ \hline
Walk & 1.00 & 0.99 & 0.97 & 0.97 & {\bf 0.94} & 0.40 & 0.53 & 0.49 & 0.37 & {\bf 0.36} & 14.25 & 10.00 & 16.00 & 20.00 & {\bf 24.25} & 0.28 & 0.28 & 0.35 & 0.41 & {\bf 0.47} \\ \hline
Greedy  & 0.99 & 1.00 & 0.99 & {\bf 0.98} & 0.99 & 0.35 & 0.35 & 0.39 & 0.24 & {\bf 0.23} & 1.25 & 1.75 & 2.50 & {\bf 4.00} & 3.07  & 0.08 & 0.04 & 0.08 & 0.12 & {\bf 0.13} \\ \hline
Info & 0.99 & 1.00 & 0.99 & 0.99 & {\bf 0.98} & 0.32 & 0.29 & 0.35 & 0.22 & {\bf 0.19} & 1.25 & 1.00 & {\bf 5.25} & 2.00 & 4.00 & 0.98 & 1.11 & 2.01 & {\bf 2.15} & 2.09 \\ \hline
Labprop & 0.85 & 0.93 & 0.96 & 0.89 & {\bf 0.79} & 0.49 & 0.44 & 0.40 & 0.38 & {\bf 0.29} & 5.00 & 3.50 & 5.00 & 7.00 & {\bf 7.20} & 0.89 & 0.95 & 1.25 & 1.50 & {\bf 1.82} \\ \hline
Eig & 0.93 & 0.97 & 0.98 & 0.92 & {\bf 0.89} & 0.21 & 0.17 & 0.11 & 0.08 & {\bf 0.07} & {\bf 5.00} & 1.25 & 1.50 & 3.00  & 3.34 & 0.68 & 0.64 & 0.70 & 0.73 & {\bf 0.75} \\ \hline

\end{tabular}}
\vspace{-5mm}
\end{table*}

\begin{table}[!h]
\centering
\caption{Accuracy of two competing community deception methods: (1) S: SADDEN (best baseline), and (2) N: NEURAL over 4 real world networks: (A) Dolphin, (B) Lesmis, (C) Polbooks, (D) Adjn, and (E) Power, such that $\beta_D = 0.3\beta$; $\beta_A = 0.7\beta$, $\beta=0.3|V_C|$}
\label{tab:add_res_extra}
\scalebox{0.9}{\begin{tabular}{|c|c|c|c|c|c|c|c|c|}

\multicolumn{9}{c}{{\bf (A) Dolphin}}\\\hline
Comm. Det. & \multicolumn{2}{c|}{NMI} & \multicolumn{2}{c|}{MNMI} & \multicolumn{2}{c|}{CommS} & \multicolumn{2}{c|}{CommU} \\ \cline{2-9}
Algo. & S & N  & S & N & S & N & S & N\\ \hline
Louv  & 0.81 & {\bf 0.79} & 0.25 & {\bf 0.23} & 2.60 & {\bf 3.00} & {\bf 0.81} & 0.68\\ \hline
Walk & 0.72 & {\bf 0.65} & 0.37 & {\bf 0.11} & 2.50 & {\bf 4.00} & 0.81 & {\bf 1.04} \\ \hline
Greedy & 0.86 & {\bf 0.78} & 0.32 & {\bf 0.27} & 2.25 & {\bf 2.75} & 0.66 & {\bf 1.11} \\ \hline
Info & 0.84 & {\bf 0.82} & {\bf 0.24} & 0.27 & 2.50 & {\bf 3.80} & 0.84 & {\bf 0.89} \\ \hline
Labprop & 0.75 & {\bf 0.69} & 0.15 & {\bf 0.10} & 1.80 & {\bf 2.50} & 0.92 & {\bf 1.56} \\ \hline
Eig & 0.81 & {\bf 0.78} & 0.22 & {\bf 0.16} & 2.00 & {\bf 2.20} & 1.04 & {\bf 1.13} \\ \hline

\multicolumn{9}{c}{{\bf (B) Lesmis}}\\\hline
Comm. Det. & \multicolumn{2}{c|}{NMI} & \multicolumn{2}{c|}{MNMI} & \multicolumn{2}{c|}{CommS} & \multicolumn{2}{c|}{CommU} \\ \cline{2-9}
Algo. & S & N  & S & N & S & N & S & N\\ \hline
Louv  & 0.89 & {\bf 0.87} & 0.33 & {\bf 0.31} & 2.13 & {\bf 2.17} & 0.83 & {\bf 1.43}\\ \hline
Walk & 0.92 & {\bf 0.90} & 0.36 & {\bf 0.33} & {\bf 1.75} & {\bf 1.75} & {\bf 1.72} & 1.58 \\ \hline
Greedy & 0.88 & {\bf 0.81} & 0.33 & {\bf 0.32} & 2.20 & {\bf 2.60} & 1.05 & {\bf 1.09} \\ \hline
Info & 0.95 & {\bf 0.90} & 0.82 & {\bf 0.13} & 2.00 & {\bf 2.50} & {\bf 1.42} & 0.98 \\ \hline
Labprop & 0.73 & {\bf 0.61} & 0.82 & {\bf 0.24} & 2.17 & {\bf 2.60} & 1.02 & {\bf 1.07} \\ \hline
Eig & 0.95 & {\bf 0.90} & 0.72 & {\bf 0.13} & 1.38 & {\bf 1.88} & 1.42 & {\bf 1.46} \\ \hline

\multicolumn{9}{c}{{\bf (C) Polbooks}}\\\hline
Comm. Det. & \multicolumn{2}{c|}{NMI} & \multicolumn{2}{c|}{MNMI} & \multicolumn{2}{c|}{CommS} & \multicolumn{2}{c|}{CommU} \\ \cline{2-9}
Algo. & S & N  & S & N & S & N & S & N\\ \hline
Louv  & 0.95 & {\bf 0.94} & 0.30 & {\bf 0.28} & {\bf 1.75} & {\bf 1.75} & 0.63 & {\bf 0.74}\\ \hline
Walk & {\bf 0.91} & 0.92 & {\bf 0.31} & {\bf 0.31} & 1.75 & {\bf 2.00} & 1.25 & {\bf 1.30} \\ \hline
Greedy & 0.92 & {\bf 0.91} & 0.32 & {\bf 0.31} & 1.75 & {\bf 2.00} & 0.76 & {\bf 0.95} \\ \hline
Info & 0.94 & {\bf 0.92} & 0.42 & {\bf 0.36} & {\bf 2.50} & 2.00 & 0.47 & {\bf 0.85} \\ \hline
Labprop & 0.89 & {\bf 0.81} & 0.35 & {\bf 0.23} & 2.00 & {\bf 2.10} & 0.23 & {\bf 0.26} \\ \hline
Eig & 0.92 & {\bf 0.90} & 0.20 & {\bf 0.18} & {\bf 1.75} & 1.50 & 1.16 & {\bf 1.80} \\ \hline

\multicolumn{9}{c}{{\bf (D) Adjn}}\\\hline
Comm. Det. & \multicolumn{2}{c|}{NMI} & \multicolumn{2}{c|}{MNMI} & \multicolumn{2}{c|}{CommS} & \multicolumn{2}{c|}{CommU} \\ \cline{2-9}
Algo. & S & N  & S & N & S & N & S & N\\ \hline
Louv  & 0.63 & {\bf 0.62} & 0.25 & {\bf 0.23} & 3.71 & {\bf 4.00} & 0.95 & {\bf 1.09}\\ \hline
Walk & 0.96 & {\bf 0.94} & 0.37 & {\bf 0.31} & {\bf 1.60} & 1.48 & 0.55 & {\bf 0.88} \\ \hline
Greedy & 0.67 & {\bf 0.56} & 0.24 & {\bf 0.22} & 3.28 & {\bf 4.00} & 1.16 & {\bf 1.18} \\ \hline
Info & 0.57 & {\bf 0.48} & {\bf 0.02} & 0.06 & 1.50 & {\bf 2.50} & 0.78 & {\bf 0.94} \\ \hline
Labprop & 1.00 & {\bf 0.00} & 1.00 & {\bf 0.00} & 1.00 & {\bf 4.00} & 0.02 & {\bf 0.15} \\ \hline
Eig & {\bf 0.68} & 0.69 & 0.28 & {\bf 0.24} & 2.00 & {\bf 2.10} & 0.57 & {\bf 1.36} \\ \hline

\multicolumn{9}{c}{{\bf (E) Power}}\\\hline
Comm. Det. & \multicolumn{2}{c|}{NMI} & \multicolumn{2}{c|}{MNMI} & \multicolumn{2}{c|}{CommS} & \multicolumn{2}{c|}{CommU} \\ \cline{2-9}
Algo. & S & N  & S & N & S & N & S & N\\ \hline
Louv  & 0.98 & {\bf 0.96} & 0.33 & {\bf 0.06} & 3.00 & {\bf 3.20} & 0.45 & {\bf 0.76}\\ \hline
Walk & 0.94 & {\bf 0.93} & 0.39 & {\bf 0.34} & 14.00 & {\bf 18.00} & 1.01 & {\bf 1.51} \\ \hline
Greedy & 0.95 & {\bf 0.92} & {\bf 0.05} & 0.10 & 3.00 & {\bf 5.00} & 0.62 & {\bf 0.89} \\ \hline
Info & 0.98 & {\bf 0.97} & 0.24 & {\bf 0.14} & 1.14 & {\bf 3.00} & {\bf 1.21} & 0.75 \\ \hline
Labprop & 0.92 & {\bf 0.89} & 0.35 & {\bf 0.26} & {\bf 6.50} & 5.00 & 0.81 & {\bf 1.02} \\ \hline
Eig & {\bf 0.81} & 0.88 & 0.12 & {\bf 0.06} & 3.50 & {\bf 4.00} & 0.61 & {\bf 0.77} \\ \hline

\end{tabular}}
\vspace{-7mm}
\end{table}

\begin{table}[!ht]
\centering
\caption{Accuracy of two competing community deception methods: (1) S: SADDEN (best baseline), and (2) N: NEURAL over 4 real world networks: (A) Dolphin, (B) Lesmis, (C) Polbooks, (D) Adjn, and (E) Power, such that $\beta_D = 0.7\beta$; $\beta_A = 0.3\beta$, $\beta=0.3|V_C|$}
\label{tab:del_res_extra}
\scalebox{0.9}{\begin{tabular}{|c|c|c|c|c|c|c|c|c|}

\multicolumn{9}{c}{{\bf (A) Dolphin}}\\\hline
Comm. Det. & \multicolumn{2}{c|}{NMI} & \multicolumn{2}{c|}{MNMI} & \multicolumn{2}{c|}{CommS} & \multicolumn{2}{c|}{CommU} \\ \cline{2-9}
Algo. & S & N  & S & N & S & N & S & N\\ \hline
Louv  & 0.91 & {\bf 0.80} & 0.22 & {\bf 0.15} & 2.40 & {\bf 3.00} & 0.67 & {\bf 0.69}\\ \hline
Walk & 0.71 & {\bf 0.66} & 0.34 & {\bf 0.09} & 3.00 & {\bf 4.00} & 0.89 & {\bf 0.97} \\ \hline
Greedy & {\bf 0.80} & 0.86 & 0.30 & {\bf 0.24} & 2.25 & {\bf 2.50} & 1.09 & {\bf 1.15} \\ \hline
Info & 0.87 & {\bf 0.85} & 0.24 & {\bf 0.23} & 2.67 & {\bf 3.00} & 1.02 & {\bf 1.14} \\ \hline
Labprop & 0.72 & {\bf 0.66} & 0.24 & {\bf 0.08} & {\bf 2.50} & 2.00 & 0.57 & {\bf 0.95} \\ \hline
Eig & {\bf 0.76} & 0.78 & 0.28 & {\bf 0.18} & 2.00 & {\bf 2.20} & 1.05 & {\bf 1.08} \\ \hline

\multicolumn{9}{c}{{\bf (B) Lesmis}}\\\hline
Comm. Det. & \multicolumn{2}{c|}{NMI} & \multicolumn{2}{c|}{MNMI} & \multicolumn{2}{c|}{CommS} & \multicolumn{2}{c|}{CommU} \\ \cline{2-9}
Algo. & S & N  & S & N & S & N & S & N\\ \hline
Louv  & 0.90 & {\bf 0.86} & 0.32 & {\bf 0.30} & {\bf 2.33} & {\bf 2.33} & 0.90 & {\bf 1.04}\\ \hline
Walk & 0.95 & {\bf 0.94} & 0.31 & {\bf 0.24} & 1.75 & {\bf 1.88} & {\bf 1.70} & 1.52 \\ \hline
Greedy & 0.84 & {\bf 0.83} & 0.32 & {\bf 0.29} & 2.20 & {\bf 2.40} & 0.96 & {\bf 1.05} \\ \hline
Info & 0.95 & {\bf 0.94} & 0.68 & {\bf 0.51} & {\bf 2.00} & {\bf 2.00} & 1.15 & {\bf 1.19} \\ \hline
Labprop & 0.75 & {\bf 0.73} & 0.68 & {\bf 0.22} & 1.20 & {\bf 2.00} & 1.09 & {\bf 1.59} \\ \hline
Eig & 0.92 & {\bf 0.91} & 0.16 & {\bf 0.15} & {\bf 1.63} & 1.34 & 0.87 & {\bf 1.62} \\ \hline

\multicolumn{9}{c}{{\bf (C) Polbooks}}\\\hline
Comm. Det. & \multicolumn{2}{c|}{NMI} & \multicolumn{2}{c|}{MNMI} & \multicolumn{2}{c|}{CommS} & \multicolumn{2}{c|}{CommU} \\ \cline{2-9}
Algo. & S & N  & S & N & S & N & S & N\\ \hline
Louv  & {\bf 0.90} & {\bf 0.90} & 0.29 & {\bf 0.27} & {\bf 2.00} & {\bf 2.00} & 0.88 & {\bf 0.89}\\ \hline
Walk & {\bf 0.90} & 0.92 & 0.35 & {\bf 0.34} & 2.00 & {\bf 2.10} & 0.66 & {\bf 0.73} \\ \hline
Greedy & 0.95 & {\bf 0.92} & 0.33 & {\bf 0.30} & 1.50 & {\bf 1.75} & 1.02 & {\bf 1.14} \\ \hline
Info & 0.96 & {\bf 0.91} & 0.38 & {\bf 0.27} & 2.17 & {\bf 2.19} & 0.81 & {\bf 0.88} \\ \hline
Labprop & 0.82 & {\bf 0.78} & 0.29 & {\bf 0.24} & 2.00 & {\bf 2.33} & 1.59 & {\bf 1.76} \\ \hline
Eig & 0.91 & {\bf 0.89} & 0.19 & {\bf 0.18} & {\bf 2.00} & 1.75 & {\bf 1.08} & 0.83 \\ \hline

\multicolumn{9}{c}{{\bf (D) Adjn}}\\\hline
Comm. Det. & \multicolumn{2}{c|}{NMI} & \multicolumn{2}{c|}{MNMI} & \multicolumn{2}{c|}{CommS} & \multicolumn{2}{c|}{CommU} \\ \cline{2-9}
Algo. & S & N  & S & N & S & N & S & N\\ \hline
Louv  & {\bf 0.54} & 0.62 & 0.23 & {\bf 0.21} & 4.00 & {\bf 4.43} & 0.99 & {\bf 1.16}\\ \hline
Walk & 0.95 & {\bf 0.94} & {\bf 0.36} & 0.38 & 1.68 & {\bf 1.69} & 0.71 & {\bf 0.84} \\ \hline
Greedy & 0.65 & {\bf 0.63} & 0.26 & {\bf 0.24} & 3.57 & {\bf 3.86} & 1.06 & {\bf 1.21} \\ \hline
Info & 0.68 & {\bf 0.56} & 0.12 & {\bf 0.05} & 1.50 & {\bf 2.50} & {\bf 0.74} & 0.37 \\ \hline
Labprop & 1.00 & {\bf 0.00} & 1.00 & {\bf 0.00} & 1.00 & {\bf 3.00} & 0.04 & {\bf 0.10} \\ \hline
Eig & {\bf 0.69} & 0.73 & 0.27 & {\bf 0.21} & 2.30 & {\bf 2.70} & 0.96 & {\bf 1.38} \\ \hline

\multicolumn{9}{c}{{\bf (E) Power}}\\\hline
Comm. Det. & \multicolumn{2}{c|}{NMI} & \multicolumn{2}{c|}{MNMI} & \multicolumn{2}{c|}{CommS} & \multicolumn{2}{c|}{CommU} \\ \cline{2-9}
Algo. & S & N  & S & N & S & N & S & N\\ \hline
Louv  & 0.98 & {\bf 0.96} & 0.34 & {\bf 0.09} & 3.50 & {\bf 5.00} & 0.65 & {\bf 1.21}\\ \hline
Walk & {\bf 0.93} & 0.96 & 0.38 & {\bf 0.34} & 10.00 & {\bf 24.00} & 1.32 & {\bf 2.44} \\ \hline
Greedy & 0.95 & {\bf 0.91} & 0.15 & {\bf 0.12} & 3.10 & {\bf 7.00} & 0.67 & {\bf 1.56} \\ \hline
Info & 0.98 & {\bf 0.96} & 0.29 & {\bf 0.12} & 1.10 & {\bf 2.00} & {\bf 1.03} & 0.55 \\ \hline
Labprop & 0.93 & {\bf 0.89} & 0.37 & {\bf 0.26} & 3.50 & {\bf 4.00} & 0.90 & {\bf 1.31} \\ \hline
Eig & {\bf 0.80} & 0.82 & 0.14 & {\bf 0.04} & 3.50 & {\bf 4.00} & 0.66 & {\bf 0.76} \\ \hline

\end{tabular}}
\vspace{-5mm}
\end{table}

\end{document}